\documentclass[aps,prl,twocolumn,superscriptaddress]{revtex4-2}
\usepackage{graphicx}  
\usepackage{dcolumn}   
\usepackage{bm}        
\usepackage{amssymb}   
\usepackage{amsmath}
\usepackage[normalem]{ulem}
\newcommand{\mbf}[1]{\mathbf{#1}}

\usepackage[svgnames]{xcolor} 

\usepackage[colorlinks=True, linkcolor=SteelBlue,
            citecolor=SteelBlue,urlcolor=SteelBlue]{hyperref}
\usepackage[capitalise]{cleveref}
\usepackage{orcidlink}
\usepackage{extpfeil}
\allowdisplaybreaks

\defcitealias{nlrl_prd}{KYV II}

\begin{document}

\title{Resonance locking of anharmonic $g$-modes in coalescing neutron star binaries}
\author{K.J. Kwon\,\orcidlink{0000-0001-9802-362X}}
\affiliation{Department of Physics, University of California, Santa Barbara, CA 93106, USA}
\author{Hang Yu\,\orcidlink{0000-0002-6011-6190}}
\affiliation{eXtreme Gravity Institute, Department of Physics, Montana State University, Bozeman, MT 59717, USA}
\author{Tejaswi Venumadhav\,\orcidlink{0000-0002-1661-2138}}
\affiliation{Department of Physics, University of California, Santa Barbara, CA 93106, USA}
\affiliation{International Centre for Theoretical Sciences, Tata Institute of Fundamental Research, Bangalore 560089, India}
\date{\today}

\begin{abstract}
Neutron stars in coalescing binaries deform due to the tidal gravitational fields generated by their companions.
During the inspiral phase, the tidal deformation is dominated by the fundamental oscillation~($f$-) mode of the stars.
The tide also has sub-dominant gravity~($g$-) modes that are resonantly excited when the linear tidal forcing sweeps through their eigenfrequencies. 
Beyond the linear order in perturbed fluid displacement, the $g$-modes are anharmonic, \emph{i.e.}, their oscillation frequencies depend on the mode energy. 
For the lowest-order $g$-mode, we show that when the tidal forcing reaches its linear eigenfrequency, the mode starts to dynamically adjust its energy so that its nonlinearly shifted oscillation frequency always matches that of the driving field. This phenomenon, which we term `resonance locking'~(RL), persists through the rest of the inspiral, and hence, the mode grows to substantially larger energies than in the linear theory. 
Using a $1.4$--$1.4\, M_{\odot}$ binary neutron star system with the SLy4 equation of state, we find this results in an extra correction to the frequency-domain gravitational wave (GW) phase of $|\Delta \Psi|\approx 3\,{\rm rad}$ accumulated from the onset of RL at the GW frequency of $94\,{\rm Hz}$ to the merger at $1.05\,{\rm kHz}$. 
This effect probes details of the internal structure of merging neutron stars beyond their bulk properties such as tidal deformability.
\end{abstract}

\maketitle

\emph{Introduction.---}Neutron stars~(NSs) are one of the most important astrophysical probes of physics at extreme densities. In a binary system containing at least one NS, the star deforms in the companion's tidal field, and this deformation impacts the gravitational wave (GW) signal emitted during the inspiral regime~\cite{Lai:94a, Lai:94b, Lai:94c, kokkotas95, mora04, flanagan08, read09, damour10, hinderer10, damour12, delPozzo13, read13, Bini:14, wade14, Bernuzzi:15, Steinhoff:16, Hinderer:16, Ma:20, Steinhoff:21,  Kuan:22, Kuan:23, Yu:24a}. 
The effect leads to an extra phase correction to the gravitational waveform since the deformation extracts energy from the orbit and modifies the rate of the inspiral. 
The dephasing is typically parameterized by a tidal deformability parameter, which is ultimately determined by the nuclear equation of state~\cite{flanagan08, damour12}.
In essence, the tidal dynamics in binary NS~(BNS) inspirals create a link between observable GW signals and the underlying nuclear physics.

The first detection of the binary neutron star merger, GW170817, yielded constraints on the tidal deformability of the components~\cite{gw170817, gw170817_improve}, which further enabled constraints on the equation of state at supranuclear densities~\cite{gw170817_eos, virgo_eos}. 
With improvements in the sensitivity of current detectors, such as the Advanced LIGO~\cite{aligo}, Virgo~\cite{virgo}, and KAGRA~\cite{kagra}, and the future commissioning of next-generation detectors~\cite{cosmic_explorer, einstein_telescope}, the number of detectable BNS inspiral signals will increase~\cite{prospect18, chan18, lenon21, mukhopadhyay24}, potentially enabling more precise constraints to be placed on nuclear equation of state. 
Therefore, accurately modeling the tidal interaction in NS binaries and its imprint on GW signals is a crucial and timely task. 

A tidally perturbed star deviates from its equilibrium configuration and the induced displacement can be viewed as a collective excitation of eigenmodes that describe stellar oscillation. In this picture, computing the time-dependent tidal deformation is equivalent to modeling the amplitudes of eigenmodes throughout inspirals. At the lowest order in hydrodynamics, the motion of the eigenmodes is linearly driven. The effect of linearly excited modes on GW phases is dominated by fundamental oscillation~($f$-) modes, which describe the large-scale, nearly adiabatic response of stars to tidal forces.
At the lower-frequency regime of the eigenmode spectrum where gravity~($g$-) modes lie, dynamical effects can become significant when modeling the mode amplitude. During the inspiral, the tidal forcing frequency evolves and crosses the natural frequencies of the spectrum of $g$-modes, which leads to their resonant excitation~\cite{Lai:94c, Yu:17a, Yu:17b, Kuan:21, Kuan:21b} in the linear tide theory. However, the impact of resonantly excited $g$-modes on the gravitational waveform is small compared to the $f$-mode's contribution \cite{Lai:94c,Yu:17b}.

Beyond the linear tide, eigenmodes are also driven by the coupling between themselves and nonlinear tidal forcing \cite{VanHoolst:94, Schenk:02}. Such nonlinear hydrodynamic interactions lead to additional complexity in the evolution of modes, hence the GW signal; Yu \emph{et al.}~\cite{Yu:23a} found that nonlinearly driven $f$-modes can lead to a cumulative phase correction of $\mathcal{O}(1)$~radian near the BNS merger on top of the dephasing due to linearly excited $f$-modes.

In this Letter, we investigate the dynamics of lowest-order $g$-modes accounting for nonlinearity. We find that the nonlinear interactions make the oscillations anharmonic~(\emph{i.e.}, causes energy-dependent shifts in mode frequencies \cite{Landau:82, Yu:21, Yu:23a}), which results in a distinctive evolution of $g$-modes compared to the linearly driven case.
In the linear picture, a $g$-mode only significantly interacts with the orbit momentarily around its resonance, but if we account for the anharmonicity associated with nonlinear evolution, the mode can be locked in a near-resonant state and stay coupled with the orbit for an extended portion of the inspiral.
We present a simple model to describe the resulting behavior of $g$-modes and discuss its impact on GW phases.

\emph{Formalism.}---Let $\bm{\chi}(\mbf{x},t)$ denote the Lagrangian displacement of the perturbed fluid at location $\mbf{x}$ in a neutron star and time $t$. We can decompose the displacement into eigenmodes $\bm{\chi}(\mbf{x},t)=\sum_a \chi_a(t) \bm{\xi}_a(\mbf{x})$, where $\chi_a$ and $\bm{\xi}_a$ are the amplitude and spatial eigenfunction of a mode $a$ (this is a configuration space decomposition; it is also possible to do it in phase space~\cite{Schenk:02}).
The eigenmodes obey orthogonality relations; we use the following normalization,
\begin{align}
&\omega_a^2\int dV \rho \, \bm{\xi}_a^\ast\cdot \bm{\xi}_b = E_0\delta_{ab},&&E_0=\frac{GM^2}{R},
\end{align}
where $\omega_a$ is the eigenfrequency of the mode, $G$ is the gravitational constant, and $M$ and $R$ are respectively the mass and radius of the deformed NS.
Then, the equation of motion for each mode reads~\cite{w12, vzh, w16, Yu:23a}
\begin{align}
    \ddot{\chi}_a + \omega_a^2 \chi_a &= \omega_a^2\left(U_a + \sum_b U_{ab}^\ast \chi_b^\ast + \sum_{bc}\kappa_{abc}\chi_b^\ast\chi_c^\ast\right.\nonumber \\
    &\phantom{=}\ \left. + \sum_{bcd}\kappa_{abcd}\chi_b^\ast\chi_c^\ast\chi_c^\ast \right),\label{eqn:eom_g}
\end{align}
where $U_a$ and $U_{ab}$ describe the linear and nonlinear tidal driving and $\kappa_{abc}$ and $\kappa_{abcd}$ are three- and four-mode coupling coefficients~\cite{w12, w16}. 

In this Letter, we focus on the $n=1$, $l=m=2$ $g$-mode, whose amplitude we denote by $\chi_g$. The driving terms that contain at least one factor of $\chi_g$ cause a nonlinear frequency shift~\cite{Yu:21, Yu:23a}, which makes the mode behave like an anharmonic oscillator~\cite{Landau:82}. Ignoring terms that are not relevant to the anharmonicity, the equation of motion of the $g$-mode can be simplified to
\begin{align}
    \ddot{\chi}_g + \omega_{\rm eff}^2\chi_g = \omega_g^2 U_g,\label{eqn:eom_g2}
\end{align}
where $\omega_{\rm eff}$ is the effective frequency of the $g$-mode,
\begin{align}
    &\omega_{\rm eff}^2=\omega_g^2(1-\kappa_{\rm eff}|\chi_g|^2 + i\gamma_{\rm eff}{\rm Im}[\chi_f^\ast \chi_g]),\label{eqn:eff_freq}\\
    &\kappa_{\rm eff} = 3\kappa_{gg^\ast gg^\ast} + 4\sum_{\beta}^{m_\beta=0}\kappa^2_{gg^\ast \beta} + 2\sum_{\beta}^{m_\beta=-4}\kappa^2_{gg\beta}<0,\label{eqn:k_eff}\\
    &\gamma_{\rm eff}=3\kappa_{gg g^\ast f^\ast} + \sum_\beta^{m_\beta=0}2\kappa_{gg^\ast \beta} \left (2\kappa_{\beta g^\ast f} + \frac{U_{\beta g^\ast }^\ast}{\chi_f^\ast}\right)\nonumber \\
    &\phantom{\gamma_{\rm eff}=}\ +\sum^{m_\beta=-4}_\beta \kappa_{gg\beta}\left(2\kappa_{gf\beta}+\frac{U_{\beta g}^\ast}{\chi_f^\ast}\right).\label{eqn:gamma_eff}
\end{align}
Here, $\chi_f$ is the amplitude of the $l=m=2$ $f$-mode and the starred subscripts denote the complex conjugate modes. In each summation, $\beta$ runs over adiabatically growing $f$- and pressure~($p$-) modes with specified azimuthal numbers~($m_\beta$). Finally, we note that the sign of $\gamma_{\rm eff}$ satisfies ${\rm Sign}[\gamma_{\rm eff}]={\rm Sign}[U_f^\ast U_g]$. For details on how we derive $\omega_{\rm eff}$, $\kappa_{\rm eff}$, and $\gamma_{\rm eff}$, we refer readers to Kwon, Yu, and Venumadhav~(hereafter \citetalias{nlrl_prd}).

In the remainder of this Letter, we demonstrate that the frequency shift due to $\kappa_{\rm eff}$ allows the $g$-mode to enter a resonance-locked state, which significantly increases the associated time and phase shifts in the GW signal. This is analogous to RL in coalescing double white dwarfs, where the frequency shift arises from the spin evolution~\cite{Fuller:12a, Burkart:13, burkart14, Yu:20a}. Additionally, although $|\kappa_{\rm eff}|\gg |\gamma_{\rm eff}|$, we show $\gamma_{\rm eff}$ counters the damping of the mode induced by GW radiation, which extends the duration of the locked phase.

\begin{figure}
    \centering
    \includegraphics[width=0.48\textwidth]{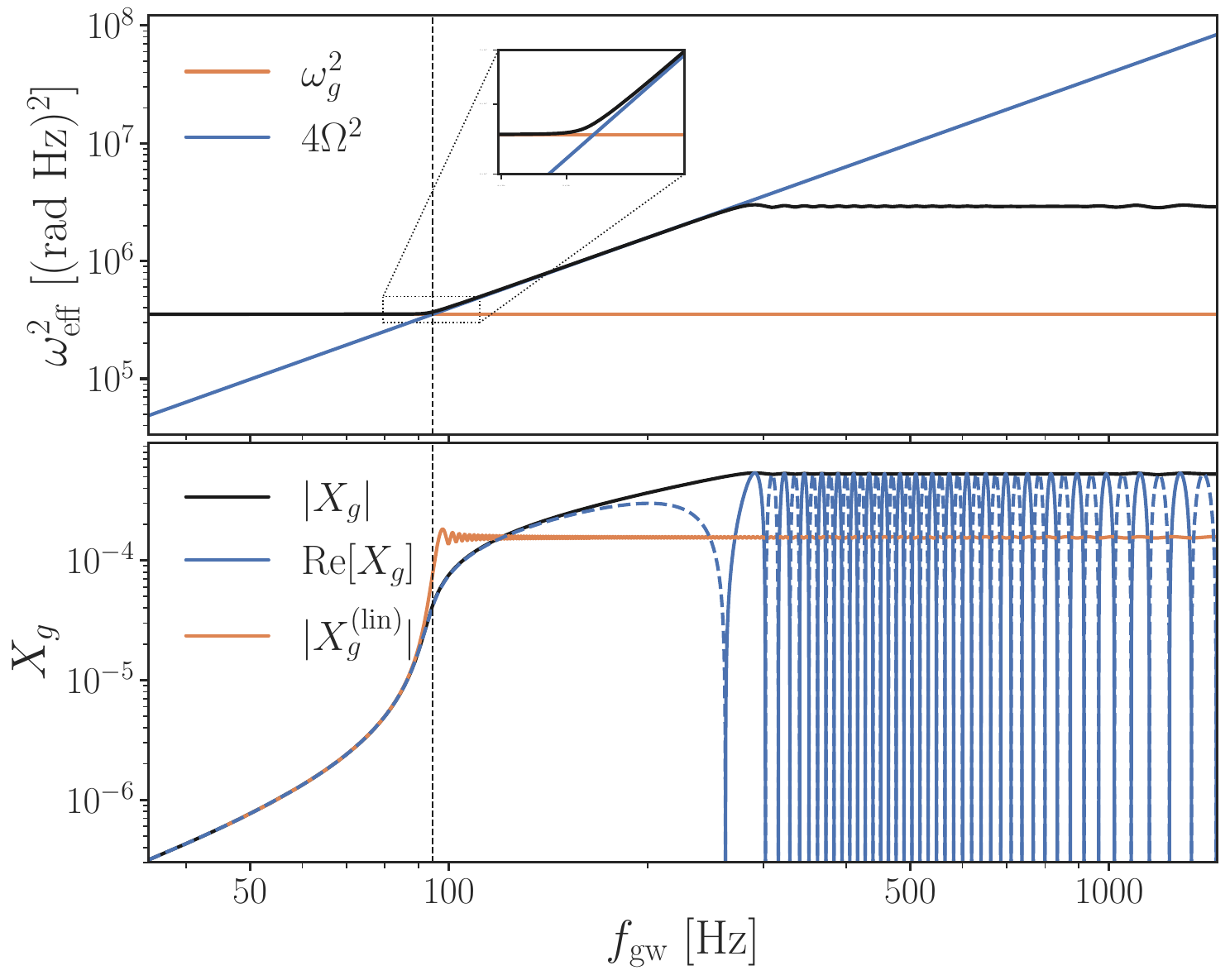} 
    
    \caption{Evolution of the $l=m=2$ $g$-mode computed using with $\kappa_{\rm eff}\approx -2\times 10^7$ given by Eq.~(\ref{eqn:k_eff}) and $\gamma_{\rm eff}=0$. \emph{Top}---effective frequency of the system (black) and the tidal forcing frequency (blue) as the binary inspirals from the initial separation. As a reference, we show the eigenfrequency of the $g$-mode in the orange horizontal line. 
    \textit{Bottom}---the mode amplitude as a function of $f_{\rm gw}$. The blue and black curves denote the magnitude and the real part of the amplitude respectively.
    } 
    \label{fig:toy_1}
\end{figure}

\begin{figure}
    \centering
    \includegraphics[width=0.48\textwidth]{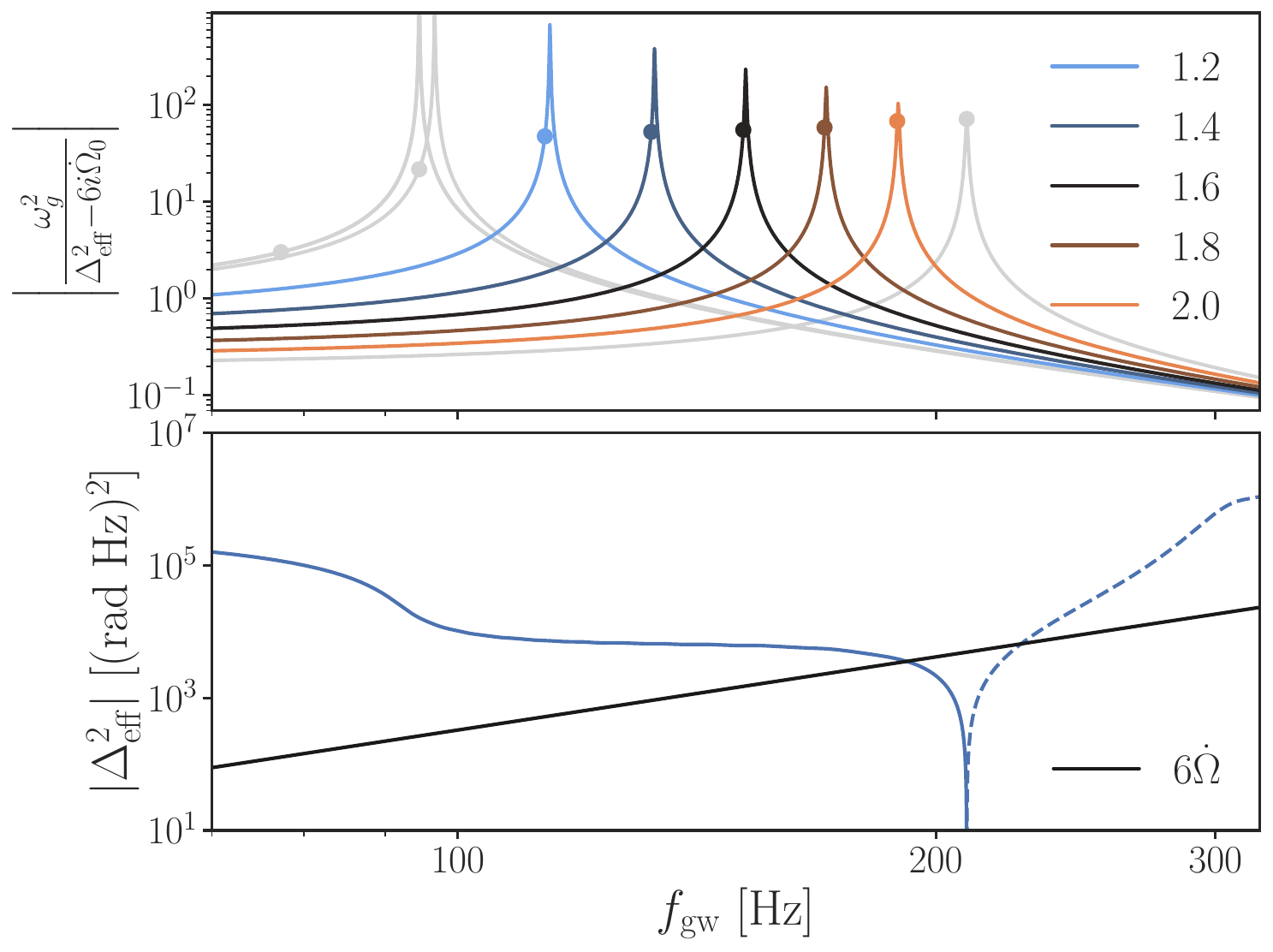}
    \caption{\emph{Top}---time-dependent Lorentzian calculated for the system shown in Fig.~\ref{fig:toy_1} at the specified points in the inspiral. We show the value of $2\Omega_0/\omega_g$ corresponding to each Lorentzian in the legend. The solid curve shows the functional shape of the Lorentzian. The `o'-markers show the actual value that Lorentzians take. 
    \emph{Bottom}---demonstration of the lock-breaking condition. We show the effective detuning (blue) and $6\dot{\Omega}$ (black). Solid and dashed curves correspond to the positive and negative signs, respectively. 
    }
    \label{fig:lor}
\end{figure}

\begin{figure}
    \centering
    \includegraphics[width=0.48\textwidth]{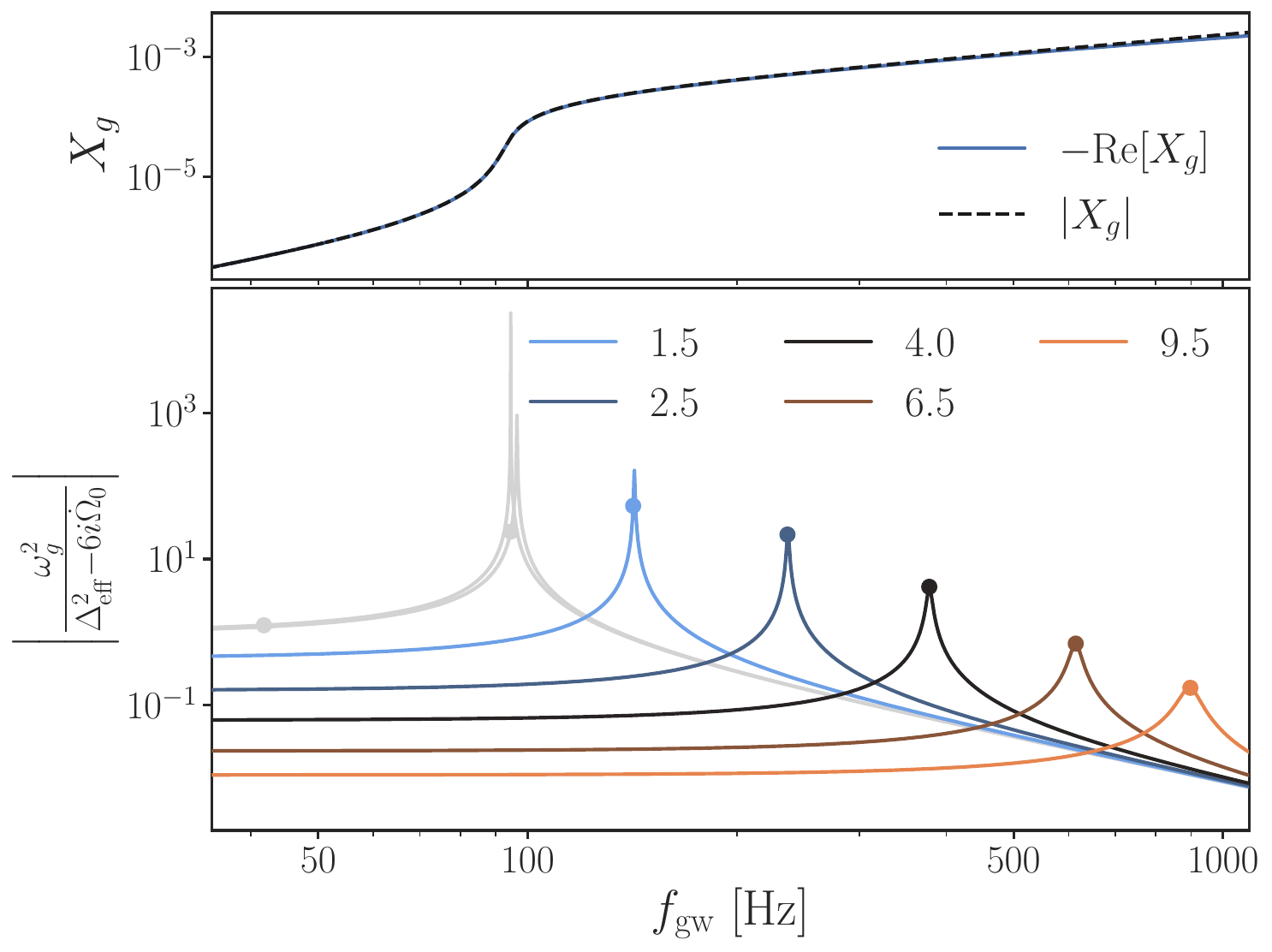}
    \caption{Evolution of the $l=m=2$ $g$-mode with $\kappa_{\rm eff}$ and $\gamma_{\rm eff}$ given by Eqs.~(\ref{eqn:k_eff})~and~(\ref{eqn:gamma_eff}). \emph{Top}---the magnitude and the real part of the $g$-mode amplitude, respectively shown in black and blue curves. 
    \emph{Bottom}---time-dependent Lorentzian at specific points in the inspiral. We show the value of $2\Omega_{0}/\omega_g$ for each point in the legend. 
    }
    \label{fig:toy_2}
\end{figure}

\begin{figure}
    \centering
    \includegraphics[width=0.48\textwidth]{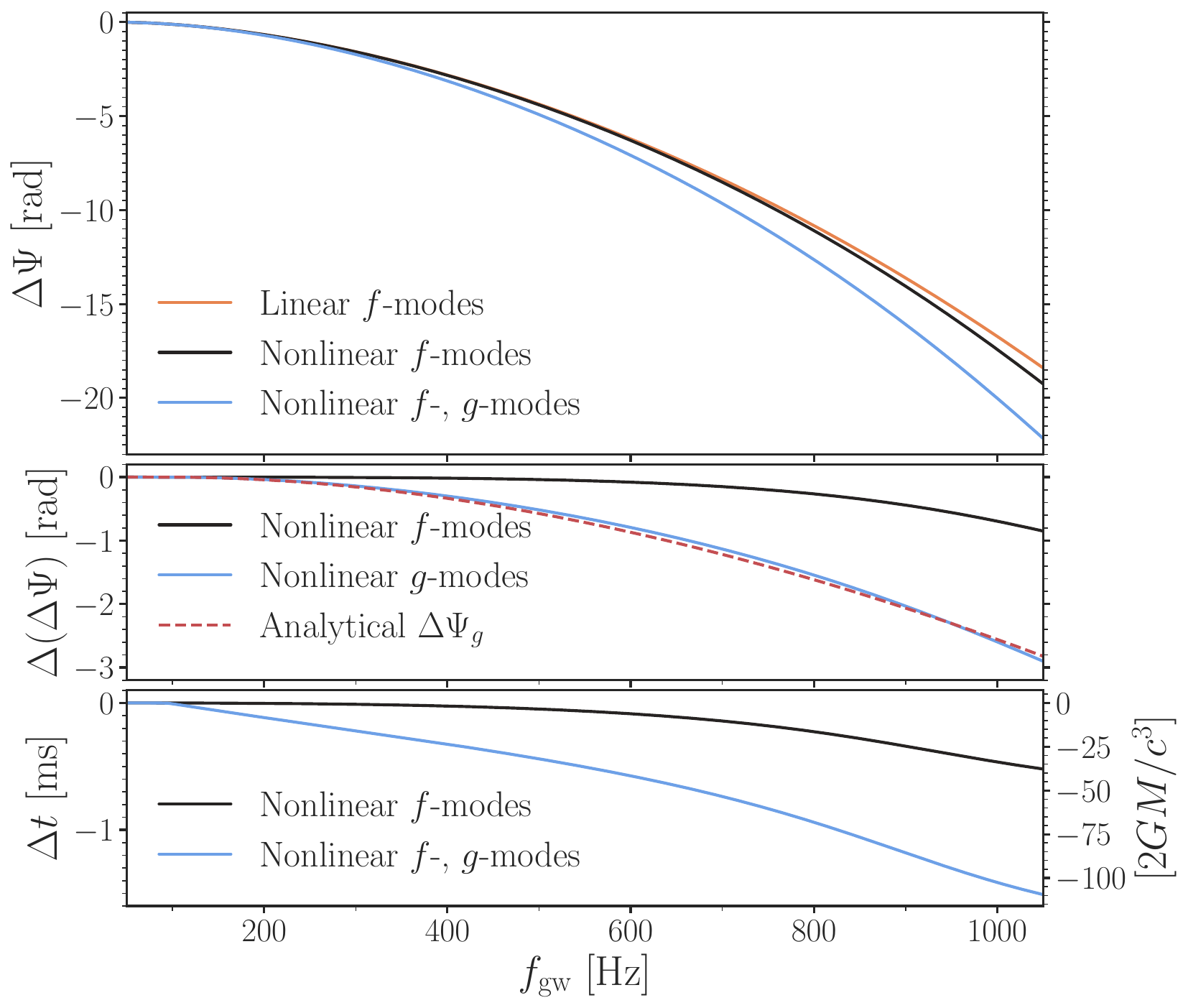}
    \caption{Tidal dephasing computed with respect to the point particle inspiraling via quadrupolar GW radiation. We evolve $l=2$ $f$- and $n=1$ $g$-modes accounting for nonlinear tidal driving, three- and four-mode coupling, as well as the cancellation due to the adiabatically sourced $p$-modes. As a reference, we include linear and nonlinear systems only with $l=2$ $f$-modes.}
    \label{fig:phase_gw}
\end{figure}

We obtain the eigenmodes and coupling coefficients by considering a non-rotating Newtonian NS governed by the SLy4 equation of state~\cite{chabanat98}. We adopt the Cowling approximation for the $g$-modes which weakly perturb gravity. The masses and radii of both the primary and companion NSs are $1.4\,M_\odot$ and $R\approx 13\,{\rm km}$. We implement the crust by setting the Brunt-V\"{a}is\"{a}l\"{a} frequency to zero when the baryon number density of the NS drops below $0.08 \,{\rm fm}^{-3}$. We compute $\kappa_{abc}$ and $\kappa_{abcd}$ following~\cite{w12, vzh, w16}. We calculate the GW radiation and its reaction at the Newtonian order but ignore the additional effect arising from the oscillating NS quadrupole~[\emph{i.e.}, Eqs.~(C26) and (C28) in Yu \emph{et al.}~\cite{Yu:23a} are ignored]. 
We neglect the tidal backreaction of modes on the orbit for the two toy models that we present in Figs.~\ref{fig:toy_1}--\ref{fig:toy_2}, but turn it back on when computing the dephasing of the GW waveform in Fig.~\ref{fig:phase_gw}.

Before proceeding, we briefly review the linear resonance of a mode which has been studied in detail~\cite{Lai:94c, Yu:24a}. 
For our $l=m=2$ $g$-mode, the resonance occurs when the frequency of tidal forcing~($2\Omega=2\dot{\Phi}=2\pi f_{\rm gw}$, where $\Omega$ is the orbital frequency, $\Phi$ is the orbital phase, and $f_{\rm gw}$ is the linear frequency of GW) reaches the eigenfrequency of a mode $\omega_g$~\cite{Lai:94c}. The mode amplitude at $2\Omega=\omega_g$, which takes half of its value obtained at $\Omega\rightarrow \infty$, can be calculated using the stationary phase approximation~\cite{Lai:94c, Yu:24a},
\begin{align}
    |\chi_g|\approx \frac{1}{2}\sqrt{\frac{\pi}{4}}\frac{\omega_g}{\dot{\Omega}^{1/2}}|U_g|\bigg|_{2\Omega=\omega_g}.\label{eqn:res_amp}
\end{align}
Once resonantly excited, the modes oscillate out of phase with respect to the orbit and their orbital backreaction can be ignored~\cite{Lai:94c}. The correction to the phase of the frequency-domain GW due to the resonant $g$-modes is $\Delta \Psi(f_{\rm gw})\sim -10^{-3} (2\pi f_{\rm gw}/\omega_{g} -1) \Theta(f_{\rm gw}-\omega_g/2\pi)\,{\rm rad}$, which is too small to be detected even for next-generation GW detectors~\cite{Lai:94c, Yu:17b}.

\emph{Resonance locking.---}When the nonlinear interactions are included, the tide due to the $g$-modes can be much stronger. 
We define $X_g=\chi_g e^{2i\Phi}$ and $V_g=U_g e^{2i\Phi}$ to factor out the rapidly oscillating orbital phase from the mode amplitude solution. 
Using a resummation technique described in Refs.~\cite{Lai:94c, Yu:24a}, $X_g$ reads
\begin{align}
&X_{g}\approx \frac{\omega_g^2 V_g}{\Delta_{\rm eff}^2-2i\Omega\left[\frac{\dot{\Omega}}{\Omega}-\frac{2}{\Delta_{\rm eff}^2}\frac{{\rm d} \Delta^2_{\rm eff}}{{\rm d}t} - 6\frac{\dot{D}}{D}\right]},\label{eqn:pre_res_amp}
\end{align}
where $\Delta_{\rm eff}^2 \equiv \omega_{\rm eff}^2-4\Omega^2$ is the effective detuning and $D$ is the orbital separation. Note that if $X_g$ is linearly and adiabatically sourced, the frequency shift~($-\omega_g^2\kappa_{\rm eff}|X_g|^2\propto \Omega^4$) in Eq.~(\ref{eqn:k_eff}) can grow faster than $-4\Omega^2$. Thus, with $\kappa_{\rm eff} < 0$, the system can evolve into a state where nonlinearly corrected $X_g$ can self-consistently satisfy
\begin{align}
    8\Omega\dot{\Omega} \gg \frac{{\rm d}\Delta_{\rm eff}^2}{{\rm d}t}\approx 0 ,\label{eqn:lock_cond}
\end{align}
and \emph{lock} $\Delta_{\rm eff}^2$ to a small and nearly constant value, which amplifies the Lorentzian $\omega_g^2/\Delta_{\rm eff}^2$ that enters Eq.~(\ref{eqn:pre_res_amp}). Additionally, $\Delta_{\rm eff}^2$ changes at a much slower rate than the detuning in the linear theory, ${\rm d}(\omega_g^2 - 4\Omega^2)/{\rm d}t=-8\Omega\dot{\Omega}$. Therefore, the near-resonantly amplified $g$-mode stays coupled to the orbit for much longer than the linearly driven $g$-modes, which only interact with the orbit significantly when $2\Omega\approx \omega_g$.

The dynamics of the locked $g$-mode can largely be understood analytically~\cite{Fuller:12a, Yu:20a}. The locking condition Eq.~(\ref{eqn:lock_cond}) requires the mode amplitude to be determined up to a constant $C$ by $\kappa_{\rm eff}\omega_g^2|X_g|^2=\omega_g^2 - 4\Omega^2 + C$. We can find $C$ by matching this locked amplitude with the equilibrium solution of $X_g$ in Eq.~(\ref{eqn:pre_res_amp}) at $2\Omega=\omega_g$, which yields,
\begin{align}
    |X_g|^2\approx -\frac{4\Omega^2}{\omega_g^2 \kappa_{\rm eff}} + \frac{1}{\kappa_{\rm eff}} + \left(\frac{V_g|_{2\Omega=\omega_g}}{\kappa_{\rm eff}}\right)^{2/3}.\label{eqn:mode_amp_locking_phase}
\end{align}
Simultaneously, $|X_g|$ must be smaller than Eq.~(\ref{eqn:res_amp}) as the mode evades the linear resonance. This defines a constraint that $\kappa_{\rm eff}$ must satisfy for the mode to enter the locked phase:
\begin{align}
    |\kappa_{\rm eff}|> \frac{1}{\omega_g^3V_g^2}\left(\frac{8\dot{\Omega}}{\pi}\right)^{3/2}\bigg|_{2\Omega=\omega_g}.
    \label{eqn:kap_crit}
\end{align}
For our NS model, we find that only the $n=1$ $g$-mode satisfies the requirement.

In Fig.~\ref{fig:toy_1}, we demonstrate this effect by evolving the $g$-mode with $\kappa_{\rm eff}\approx -2\times 10^7$ given by Eq.~(\ref{eqn:k_eff}) and $\gamma_{\rm eff}=0$. The top panel shows that $\omega_{\rm eff}$~(black curve) as a function of $f_{\rm gw}$. We find that $\omega_{\rm eff}$ tracks the tidal forcing frequency~(blue curve) as the binary inspirals beyond the point of linear resonance where $2\Omega=\omega_g$~(vertical black dashed line).
For reference, we show the $g$-mode eigenfrequency $\omega_g/2\pi\approx94\, {\rm Hz}$ using the orange line. The tidal forcing frequency eventually crosses $\omega_{\rm eff}$ at $f_{\rm gw}\approx 280\, {\rm Hz}$, after which we expect the lock to break for a reason we explain later.

We verify this in the second panel where we show $|X_g|$ in the linear and nonlinear systems~(orange and black, respectively). For the nonlinear case, we additionally show ${\rm Re}[X_g]$ (blue), where the solid and dashed curves indicate the positive and negative signs. The nonlinear system oscillates resonantly past $f_{\rm gw}\approx 200\,{\rm Hz}$.

The lock can break due to the damping induced by GW radiation. Eq.~(\ref{eqn:mode_amp_locking_phase}) suggests $|X_g|^2\sim -4\Omega^2/\omega_g^2 \kappa_{\rm eff}$ asymptotically as $\Omega$ increases. This in turn implies from Eq.~(\ref{eqn:pre_res_amp}) that $\Delta_{\rm eff}^2$ grows proportionally to $\Omega$.
This lets us simplify Eq.~(\ref{eqn:pre_res_amp}) as
\begin{align}
    X_g&\approx \frac{\omega_g^2 V_g}{\Delta_{\rm eff}^2-6i\dot{\Omega}}\approx 
    \frac{\omega_g^2V_g}{{\rm Re}[\Delta_{\rm eff}^2]}\left( 1 + i\frac{6\dot{\Omega} - {\rm Im}[\Delta_{\rm eff}^2]  }{{\rm Re}[\Delta_{\rm eff}^2]}\right),\label{eqn:taylor}
\end{align}
where we have used $\dot{D}/D=-2\dot{\Omega}/3\Omega$ to obtain the first equality. We find that the lock can be maintained only if the real part of  $\Delta_{\rm eff}^2$ dominates over the damping, which corresponds to the imaginary part of the denominator, \emph{i.e.},
\begin{align}
    |{\rm Re}[\Delta_{\rm eff}^2]|\gg |6\dot{\Omega} - \omega_g^2\gamma_{\rm eff}{\rm Im}[X_f^\ast X_g ]|,\label{eqn:break_cond}
\end{align}
where $\omega_g^2 \gamma_{\rm eff}{\rm Im}[\chi_f^\ast\chi_g]={\rm Im}[\Delta_{\rm eff}^2]$ [Eq.~(\ref{eqn:eff_freq})]. If the above inequality holds, the mode can self-consistently adjust $\Delta_{\rm eff}^2$ to maintain the evolution described in Eq.~(\ref{eqn:mode_amp_locking_phase}). However, if $\gamma_{\rm eff}=0$, the steeply growing inspiral-driven damping term $\dot{\Omega}\propto \Omega^{11/3}$ can eventually dominate over the detuning, hence breaking the lock. 

In Fig.~\ref{fig:lor}, we show the Lorentzian for the system shown in Fig.~\ref{fig:toy_1}. For different points in the inspiral specified by $\Omega=\Omega_0$, we show the functional form of $\omega_g^2/[\Delta_{\rm eff}^2(\Omega)-6i\dot{\Omega}_0]$~(solid curves). We mark the actual value that each Lorentzian takes with the `o'-marker. 
The anharmonicity keeps the detuning small and near-resonantly amplifies $X_g$ until Eq.~(\ref{eqn:break_cond}) no longer holds at $f_{\rm gw}\approx 200\,{\rm Hz}$.

However, with $\gamma_{\rm eff}\neq 0 $ and 
$\gamma_{\rm eff} \omega_g^2 {\rm Im}[X_f^\ast X_g]>0$, the lock can be maintained further since it can cancel the inspiral-induced damping $\propto 6\dot{\Omega}$.
We illustrate this in Fig.~\ref{fig:toy_2}, where we show the $g$-mode evolved while taking $\gamma_{\rm eff}$ into account as given by Eq.~(\ref{eqn:gamma_eff}). Using $X_g$ evaluated with $\gamma_{\rm eff}=0$ and the linear adiabatic $X_f$, ${\rm Im}[\Delta_{\rm eff}^2]\propto \Omega^{17/3}$ grows faster than $\dot{\Omega}$. Therefore, similar to the locking of the real detuning, the damping can also evolve into a locking state with 
${\rm Im}[\Delta_{\rm eff}^2]\approx 6\dot{\Omega}$. Taking similar steps leading to Eq.~(\ref{eqn:mode_amp_locking_phase}), we find $|{\rm Im}[X_g]|\simeq | 6\dot{\Omega} - {\rm Im}[\Delta_{\rm eff}^2]|\propto \Omega^{5/3}$, evolving much slower than $\dot{\Omega}\propto\Omega^{11/3}$ (see \citetalias{nlrl_prd} for a more detailed discussion).

\emph{Dephasing estimate.---}We estimate the impact of the locked $g$-modes on the frequency-domain GW phase including the nonlinear tidal driving, three- and four-mode couplings. We evolve $l=2$, $n=1$ $g$-modes along with the $l=2$ $f$-modes from an initial separation of $20R$ until contact. We calculate $\kappa_{\rm eff}$ and $\gamma_{\rm eff}$ accounting for $l=0,2,4$, adiabatic $p$-modes with eigenfrequencies smaller than the acoustic cutoff frequency of the NS. This amounts to including $p$-modes with radial order $n\leq 60$. We account for the corresponding backreaction onto the $f$-modes as well. 
As we justify in \citetalias{nlrl_prd}, we do not include the higher-order $g$-modes since they do not meet the condition in Eq.~(\ref{eqn:kap_crit}).

We calculate the dephasing of the frequency-domain waveform~($\Psi$) with respect to the waveform of a point particle inspiral. 
We use the analytical expression for the frequency-domain phase~($\Psi_{\rm pp}$) of the waveform in the point particle case under the quadrupole approximation.

The top panel of Fig.~\ref{fig:phase_gw} shows the phase correction, $\Delta \Psi \equiv \Psi - \Psi_{\rm pp}$, due to the locked $g$-modes (blue). For comparison, we include $\Delta \Psi$ due to linear~(orange) and nonlinear~(black) $f$-mode tides.  
The middle panel shows the extra phase correction on top of that due to the linear $f$-modes, caused by the nonlinear $f$- and $g$-modes~(black and blue, respectively). 
We obtained the $g$-mode contribution by subtracting $\Delta \Psi$ due to nonlinear $f$-modes alone from that due to both nonlinear $f$- and $g$-modes. 
The locked $g$-mode causes an extra phase correction of approximately~$3 \, {\rm rad}$ on top of the nonlinear and linear $f$-mode tides at $f_{\rm gw}=1.05\,{\rm kHz}$.

Using Eq.~(\ref{eqn:mode_amp_locking_phase}) and following the steps explained in \citetalias{nlrl_prd}, we analytically derive an approximation to the phase correction due to the resonantly locked $g$-modes as
\begin{align}
&\Delta \Psi_g \approx K[54a^5-5(21 + 10 F_0 -F_0^2)a^3+96+140 F_0\nonumber \\
 &\phantom{\Delta\Psi_g\approx K}- 5 F_0^2-45(1  +2F_0 )a^{-1}]\Theta(a-1),\label{eqn:deph_psi}\\
 &K=\frac{1}{2^{8/3}\kappa_{\rm eff}\eta^2}\frac{E_0}{M_{\rm t} c^2}\bigg(\frac{G M_{\rm t}\omega_g}{c^3}\bigg)^{-7/3}\\
&F_0 =-\bigg|\frac{3\pi I_g^2\kappa_{\rm eff}}{10}\bigg|^{1/3}\bigg(\frac{\omega^2_gR^3}{4GM_{\rm t}}\bigg)^{2/3}, a = \left(\frac{2\pi f_{\rm gw}}{\omega_g}\right)^{1/3},
\end{align}
where $c$ is the speed of light, $\eta$ is the symmetric mass ratio, $M_{\rm t}$ is the total mass, and $I_g$ is the overlap integral of the $g$-mode. 
Note that Eq.~(\ref{eqn:deph_psi}) accounts for the RL in a primary NS. 
The contribution from the companion NS may be obtained using the same equation, but with $E_0$, $\kappa_{\rm eff}$, $I_g$, and $\omega_g$ appropriately replaced for the companion.
We show $\Delta \Psi_g$ using a red dashed line, which agrees well with the numerically calculated dephasing for most of the inspiral.
The third panel shows that the time shift due to the nonlinear $f$- and $g$-modes relative to the linear $f$-mode tide accumulates to $\mathcal{O}(1.5)\,{\rm ms}$ toward the merger.

\emph{Discussion.}---We have demonstrated that the lowest-order $g$-mode can have a significantly larger impact on GW signals than what is predicted by linear theory \cite{Lai:94c, Yu:17a, Yu:17b, Kuan:21, Kuan:21b}, since the mode's nonlinear frequency shift enables it to be resonantly locked to the tidal forcing frequency.
The RL of $g$-modes we have presented is a generic feature of BNS inspirals, as our scaling analysis with respect to the orbital frequency implies that the real and imaginary parts of $\Delta_{\rm eff}^2$ are locked for a wide range of $\kappa_{\rm eff}$ and $\gamma_{\rm eff}$. 
This suggests that the phenomenon should be robust to different nuclear equations of state. 
The lock initiates as the gravitational wave frequency crosses the $g$-mode's natural frequency at $f_{\rm gw}\approx 94\,{\rm Hz}$, which implies that this phenomenon occurs well outside the coverage of current numerical relativity simulations~\cite{hotokezaka13, Hotokezaka:15, Dietrich:18, Foucart:19} that only include the last few orbits of the inspiral.
Therefore, our results strongly motivate further analytical studies of the effect's impact on BNS inspirals, and considerations of how to test it via numerical simulations.

Our analysis assumes that the NS is not rotating. However, the impact of spin on our analysis remains to be investigated, given that non-zero spin can greatly amplify the $g$-mode~(or inertial-gravity modes with rotation) response under the linear theory~\cite{Ho:99, Xu:17}. Furthermore, the cores of our NSs are composed of normal fluid, although they are expected to be in the superfluid state in reality~\cite{Yakovlev:99}.
The spectrum of $g$-modes of a superfluid NS can largely differ from its normal-fluid counterpart---\emph{e.g.}, the first $g$-mode is excited at a much higher frequency $\omega_g/\pi\sim300$--$500\,{\rm Hz}$ \cite{Kantor:14, Passamonti:16, Yu:17a, Yu:17b}. Thus, the impact of superfluidity on RL could be important to consider in future studies.

We have also adopted a simple prescription for the NS crust by assuming that it is fluid with neutral buoyancy and ignored any potential impacts on the core $g$-modes from crust shattering~\cite{Tsang:12, Passamonti:21, Kuan:21} or melting \cite{Pan:20}. 
Incorporating crustal dynamics into the picture could be an interesting direction for future investigations. 

The nonlinear interactions we have included are conservative and ignore fluid damping. Additionally, we have not considered the potential saturation of the $g$-modes. The standard wave-breaking condition where the shear exceeds unity~(which closely ties to the parametric instability of the $g$-modes~\cite{Weinberg:12}) 
should not alter our analysis because the GW-driven orbital decay is much faster than the rate at which parametric instability can develop~(see Weinberg \emph{et al.}~\cite{Weinberg:13}, although resonantly locked $g$-modes were not considered here). Nonetheless, whether the RL can be regulated by or interplays with other effects, such as the nonlinear $p$-$g$~instability~\cite{Weinberg:13, Venumadhav:14, w16}, is an interesting question. 

Finally, we note that our calculation is Newtonian in nature; the dynamics of RL most crucially depend on a single parameter~$\kappa_{\rm eff}$. 
Extending our analysis to post-Newtonian orders, and computing the relativistic equivalent to $\kappa_{\rm eff}$ will be an important step forward.

The authors appreciate Eliot Quataert, Lars Bildsten, and Christopher Hirata for useful discussions during the preparation of this work. 
KJK has received support from the National Science Foundation~(NSF) under Grant No. 2012086.
HY acknowledges support from NSF grant No. PHY-2308415 and from Montana NASA EPSCoR Research Infrastructure Development under award No. 80NSSC22M0042.
TV acknowledges support from NSF grants 2012086 and 2309360, the Alfred P. Sloan Foundation through grant number FG-2023-20470, the BSF through award number 2022136, and the Hellman Family Faculty Fellowship.

\bibliographystyle{apsrev}
\bibliography{prl}

\begin{thebibliography}{68}
\expandafter\ifx\csname natexlab\endcsname\relax\def\natexlab#1{#1}\fi
\expandafter\ifx\csname bibnamefont\endcsname\relax
  \def\bibnamefont#1{#1}\fi
\expandafter\ifx\csname bibfnamefont\endcsname\relax
  \def\bibfnamefont#1{#1}\fi
\expandafter\ifx\csname citenamefont\endcsname\relax
  \def\citenamefont#1{#1}\fi
\expandafter\ifx\csname url\endcsname\relax
  \def\url#1{\texttt{#1}}\fi
\expandafter\ifx\csname urlprefix\endcsname\relax\def\urlprefix{URL }\fi
\providecommand{\bibinfo}[2]{#2}
\providecommand{\eprint}[2][]{\url{#2}}

\bibitem[{\citenamefont{{Lai} et~al.}(1994{\natexlab{a}})\citenamefont{{Lai}, {Rasio}, and {Shapiro}}}]{Lai:94a}
\bibinfo{author}{\bibfnamefont{D.}~\bibnamefont{{Lai}}}, \bibinfo{author}{\bibfnamefont{F.~A.} \bibnamefont{{Rasio}}}, \bibnamefont{and} \bibinfo{author}{\bibfnamefont{S.~L.} \bibnamefont{{Shapiro}}}, \bibinfo{journal}{\apj} \textbf{\bibinfo{volume}{420}}, \bibinfo{pages}{811} (\bibinfo{year}{1994}{\natexlab{a}}), \eprint{astro-ph/9304027}.

\bibitem[{\citenamefont{{Lai} et~al.}(1994{\natexlab{b}})\citenamefont{{Lai}, {Rasio}, and {Shapiro}}}]{Lai:94b}
\bibinfo{author}{\bibfnamefont{D.}~\bibnamefont{{Lai}}}, \bibinfo{author}{\bibfnamefont{F.~A.} \bibnamefont{{Rasio}}}, \bibnamefont{and} \bibinfo{author}{\bibfnamefont{S.~L.} \bibnamefont{{Shapiro}}}, \bibinfo{journal}{\apj} \textbf{\bibinfo{volume}{423}}, \bibinfo{pages}{344} (\bibinfo{year}{1994}{\natexlab{b}}), \eprint{astro-ph/9307032}.

\bibitem[{\citenamefont{{Lai}}(1994)}]{Lai:94c}
\bibinfo{author}{\bibfnamefont{D.}~\bibnamefont{{Lai}}}, \bibinfo{journal}{\mnras} \textbf{\bibinfo{volume}{270}}, \bibinfo{pages}{611} (\bibinfo{year}{1994}), \eprint{astro-ph/9404062}.

\bibitem[{\citenamefont{{Kokkotas} and {Schafer}}(1995)}]{kokkotas95}
\bibinfo{author}{\bibfnamefont{K.~D.} \bibnamefont{{Kokkotas}}} \bibnamefont{and} \bibinfo{author}{\bibfnamefont{G.}~\bibnamefont{{Schafer}}}, \bibinfo{journal}{\mnras} \textbf{\bibinfo{volume}{275}}, \bibinfo{pages}{301} (\bibinfo{year}{1995}), \eprint{gr-qc/9502034}.

\bibitem[{\citenamefont{{Mora} and {Will}}(2004)}]{mora04}
\bibinfo{author}{\bibfnamefont{T.}~\bibnamefont{{Mora}}} \bibnamefont{and} \bibinfo{author}{\bibfnamefont{C.~M.} \bibnamefont{{Will}}}, \bibinfo{journal}{\prd} \textbf{\bibinfo{volume}{69}}, \bibinfo{eid}{104021} (\bibinfo{year}{2004}), \eprint{gr-qc/0312082}.

\bibitem[{\citenamefont{{Flanagan} and {Hinderer}}(2008)}]{flanagan08}
\bibinfo{author}{\bibfnamefont{{\'E}.~{\'E}.} \bibnamefont{{Flanagan}}} \bibnamefont{and} \bibinfo{author}{\bibfnamefont{T.}~\bibnamefont{{Hinderer}}}, \bibinfo{journal}{\prd} \textbf{\bibinfo{volume}{77}}, \bibinfo{eid}{021502} (\bibinfo{year}{2008}), \eprint{0709.1915}.

\bibitem[{\citenamefont{{Read} et~al.}(2009)\citenamefont{{Read}, {Markakis}, {Shibata}, {Ury{\={u}}}, {Creighton}, and {Friedman}}}]{read09}
\bibinfo{author}{\bibfnamefont{J.~S.} \bibnamefont{{Read}}}, \bibinfo{author}{\bibfnamefont{C.}~\bibnamefont{{Markakis}}}, \bibinfo{author}{\bibfnamefont{M.}~\bibnamefont{{Shibata}}}, \bibinfo{author}{\bibfnamefont{K.}~\bibnamefont{{Ury{\={u}}}}}, \bibinfo{author}{\bibfnamefont{J.~D.~E.} \bibnamefont{{Creighton}}}, \bibnamefont{and} \bibinfo{author}{\bibfnamefont{J.~L.} \bibnamefont{{Friedman}}}, \bibinfo{journal}{\prd} \textbf{\bibinfo{volume}{79}}, \bibinfo{eid}{124033} (\bibinfo{year}{2009}), \eprint{0901.3258}.

\bibitem[{\citenamefont{Damour and Nagar}(2010)}]{damour10}
\bibinfo{author}{\bibfnamefont{T.}~\bibnamefont{Damour}} \bibnamefont{and} \bibinfo{author}{\bibfnamefont{A.}~\bibnamefont{Nagar}}, \bibinfo{journal}{Phys. Rev. D} \textbf{\bibinfo{volume}{81}}, \bibinfo{pages}{084016} (\bibinfo{year}{2010}), \urlprefix\url{https://link.aps.org/doi/10.1103/PhysRevD.81.084016}.

\bibitem[{\citenamefont{Hinderer et~al.}(2010)\citenamefont{Hinderer, Lackey, Lang, and Read}}]{hinderer10}
\bibinfo{author}{\bibfnamefont{T.}~\bibnamefont{Hinderer}}, \bibinfo{author}{\bibfnamefont{B.~D.} \bibnamefont{Lackey}}, \bibinfo{author}{\bibfnamefont{R.~N.} \bibnamefont{Lang}}, \bibnamefont{and} \bibinfo{author}{\bibfnamefont{J.~S.} \bibnamefont{Read}}, \bibinfo{journal}{Phys. Rev. D} \textbf{\bibinfo{volume}{81}}, \bibinfo{pages}{123016} (\bibinfo{year}{2010}), \urlprefix\url{https://link.aps.org/doi/10.1103/PhysRevD.81.123016}.

\bibitem[{\citenamefont{{Damour} et~al.}(2012)\citenamefont{{Damour}, {Nagar}, and {Villain}}}]{damour12}
\bibinfo{author}{\bibfnamefont{T.}~\bibnamefont{{Damour}}}, \bibinfo{author}{\bibfnamefont{A.}~\bibnamefont{{Nagar}}}, \bibnamefont{and} \bibinfo{author}{\bibfnamefont{L.}~\bibnamefont{{Villain}}}, \bibinfo{journal}{\prd} \textbf{\bibinfo{volume}{85}}, \bibinfo{eid}{123007} (\bibinfo{year}{2012}), \eprint{1203.4352}.

\bibitem[{\citenamefont{Del~Pozzo et~al.}(2013)\citenamefont{Del~Pozzo, Li, Agathos, Van Den~Broeck, and Vitale}}]{delPozzo13}
\bibinfo{author}{\bibfnamefont{W.}~\bibnamefont{Del~Pozzo}}, \bibinfo{author}{\bibfnamefont{T.~G.~F.} \bibnamefont{Li}}, \bibinfo{author}{\bibfnamefont{M.}~\bibnamefont{Agathos}}, \bibinfo{author}{\bibfnamefont{C.}~\bibnamefont{Van Den~Broeck}}, \bibnamefont{and} \bibinfo{author}{\bibfnamefont{S.}~\bibnamefont{Vitale}}, \bibinfo{journal}{Phys. Rev. Lett.} \textbf{\bibinfo{volume}{111}}, \bibinfo{pages}{071101} (\bibinfo{year}{2013}), \urlprefix\url{https://link.aps.org/doi/10.1103/PhysRevLett.111.071101}.

\bibitem[{\citenamefont{Read et~al.}(2013)\citenamefont{Read, Baiotti, Creighton, Friedman, Giacomazzo, Kyutoku, Markakis, Rezzolla, Shibata, and Taniguchi}}]{read13}
\bibinfo{author}{\bibfnamefont{J.~S.} \bibnamefont{Read}}, \bibinfo{author}{\bibfnamefont{L.}~\bibnamefont{Baiotti}}, \bibinfo{author}{\bibfnamefont{J.~D.~E.} \bibnamefont{Creighton}}, \bibinfo{author}{\bibfnamefont{J.~L.} \bibnamefont{Friedman}}, \bibinfo{author}{\bibfnamefont{B.}~\bibnamefont{Giacomazzo}}, \bibinfo{author}{\bibfnamefont{K.}~\bibnamefont{Kyutoku}}, \bibinfo{author}{\bibfnamefont{C.}~\bibnamefont{Markakis}}, \bibinfo{author}{\bibfnamefont{L.}~\bibnamefont{Rezzolla}}, \bibinfo{author}{\bibfnamefont{M.}~\bibnamefont{Shibata}}, \bibnamefont{and} \bibinfo{author}{\bibfnamefont{K.}~\bibnamefont{Taniguchi}}, \bibinfo{journal}{Phys. Rev. D} \textbf{\bibinfo{volume}{88}}, \bibinfo{pages}{044042} (\bibinfo{year}{2013}), \urlprefix\url{https://link.aps.org/doi/10.1103/PhysRevD.88.044042}.

\bibitem[{\citenamefont{{Bini} and {Damour}}(2014)}]{Bini:14}
\bibinfo{author}{\bibfnamefont{D.}~\bibnamefont{{Bini}}} \bibnamefont{and} \bibinfo{author}{\bibfnamefont{T.}~\bibnamefont{{Damour}}}, \bibinfo{journal}{\prd} \textbf{\bibinfo{volume}{90}}, \bibinfo{eid}{124037} (\bibinfo{year}{2014}), \eprint{1409.6933}.

\bibitem[{\citenamefont{Wade et~al.}(2014)\citenamefont{Wade, Creighton, Ochsner, Lackey, Farr, Littenberg, and Raymond}}]{wade14}
\bibinfo{author}{\bibfnamefont{L.}~\bibnamefont{Wade}}, \bibinfo{author}{\bibfnamefont{J.~D.~E.} \bibnamefont{Creighton}}, \bibinfo{author}{\bibfnamefont{E.}~\bibnamefont{Ochsner}}, \bibinfo{author}{\bibfnamefont{B.~D.} \bibnamefont{Lackey}}, \bibinfo{author}{\bibfnamefont{B.~F.} \bibnamefont{Farr}}, \bibinfo{author}{\bibfnamefont{T.~B.} \bibnamefont{Littenberg}}, \bibnamefont{and} \bibinfo{author}{\bibfnamefont{V.}~\bibnamefont{Raymond}}, \bibinfo{journal}{Phys. Rev. D} \textbf{\bibinfo{volume}{89}}, \bibinfo{pages}{103012} (\bibinfo{year}{2014}), \urlprefix\url{https://link.aps.org/doi/10.1103/PhysRevD.89.103012}.

\bibitem[{\citenamefont{{Bernuzzi} et~al.}(2015)\citenamefont{{Bernuzzi}, {Nagar}, {Dietrich}, and {Damour}}}]{Bernuzzi:15}
\bibinfo{author}{\bibfnamefont{S.}~\bibnamefont{{Bernuzzi}}}, \bibinfo{author}{\bibfnamefont{A.}~\bibnamefont{{Nagar}}}, \bibinfo{author}{\bibfnamefont{T.}~\bibnamefont{{Dietrich}}}, \bibnamefont{and} \bibinfo{author}{\bibfnamefont{T.}~\bibnamefont{{Damour}}}, \bibinfo{journal}{\prl} \textbf{\bibinfo{volume}{114}}, \bibinfo{eid}{161103} (\bibinfo{year}{2015}), \eprint{1412.4553}.

\bibitem[{\citenamefont{{Steinhoff} et~al.}(2016)\citenamefont{{Steinhoff}, {Hinderer}, {Buonanno}, and {Taracchini}}}]{Steinhoff:16}
\bibinfo{author}{\bibfnamefont{J.}~\bibnamefont{{Steinhoff}}}, \bibinfo{author}{\bibfnamefont{T.}~\bibnamefont{{Hinderer}}}, \bibinfo{author}{\bibfnamefont{A.}~\bibnamefont{{Buonanno}}}, \bibnamefont{and} \bibinfo{author}{\bibfnamefont{A.}~\bibnamefont{{Taracchini}}}, \bibinfo{journal}{\prd} \textbf{\bibinfo{volume}{94}}, \bibinfo{eid}{104028} (\bibinfo{year}{2016}), \eprint{1608.01907}.

\bibitem[{\citenamefont{{Hinderer} et~al.}(2016)\citenamefont{{Hinderer}, {Taracchini}, {Foucart}, {Buonanno}, {Steinhoff}, {Duez}, {Kidder}, {Pfeiffer}, {Scheel}, {Szilagyi} et~al.}}]{Hinderer:16}
\bibinfo{author}{\bibfnamefont{T.}~\bibnamefont{{Hinderer}}}, \bibinfo{author}{\bibfnamefont{A.}~\bibnamefont{{Taracchini}}}, \bibinfo{author}{\bibfnamefont{F.}~\bibnamefont{{Foucart}}}, \bibinfo{author}{\bibfnamefont{A.}~\bibnamefont{{Buonanno}}}, \bibinfo{author}{\bibfnamefont{J.}~\bibnamefont{{Steinhoff}}}, \bibinfo{author}{\bibfnamefont{M.}~\bibnamefont{{Duez}}}, \bibinfo{author}{\bibfnamefont{L.~E.} \bibnamefont{{Kidder}}}, \bibinfo{author}{\bibfnamefont{H.~P.} \bibnamefont{{Pfeiffer}}}, \bibinfo{author}{\bibfnamefont{M.~A.} \bibnamefont{{Scheel}}}, \bibinfo{author}{\bibfnamefont{B.}~\bibnamefont{{Szilagyi}}}, \bibnamefont{et~al.}, \bibinfo{journal}{\prl} \textbf{\bibinfo{volume}{116}}, \bibinfo{eid}{181101} (\bibinfo{year}{2016}), \eprint{1602.00599}.

\bibitem[{\citenamefont{{Ma} et~al.}(2020)\citenamefont{{Ma}, {Yu}, and {Chen}}}]{Ma:20}
\bibinfo{author}{\bibfnamefont{S.}~\bibnamefont{{Ma}}}, \bibinfo{author}{\bibfnamefont{H.}~\bibnamefont{{Yu}}}, \bibnamefont{and} \bibinfo{author}{\bibfnamefont{Y.}~\bibnamefont{{Chen}}}, \bibinfo{journal}{\prd} \textbf{\bibinfo{volume}{101}}, \bibinfo{eid}{123020} (\bibinfo{year}{2020}), \eprint{2003.02373}.

\bibitem[{\citenamefont{{Steinhoff} et~al.}(2021)\citenamefont{{Steinhoff}, {Hinderer}, {Dietrich}, and {Foucart}}}]{Steinhoff:21}
\bibinfo{author}{\bibfnamefont{J.}~\bibnamefont{{Steinhoff}}}, \bibinfo{author}{\bibfnamefont{T.}~\bibnamefont{{Hinderer}}}, \bibinfo{author}{\bibfnamefont{T.}~\bibnamefont{{Dietrich}}}, \bibnamefont{and} \bibinfo{author}{\bibfnamefont{F.}~\bibnamefont{{Foucart}}}, \bibinfo{journal}{Physical Review Research} \textbf{\bibinfo{volume}{3}}, \bibinfo{eid}{033129} (\bibinfo{year}{2021}), \eprint{2103.06100}.

\bibitem[{\citenamefont{{Kuan} and {Kokkotas}}(2022)}]{Kuan:22}
\bibinfo{author}{\bibfnamefont{H.-J.} \bibnamefont{{Kuan}}} \bibnamefont{and} \bibinfo{author}{\bibfnamefont{K.~D.} \bibnamefont{{Kokkotas}}}, \bibinfo{journal}{\prd} \textbf{\bibinfo{volume}{106}}, \bibinfo{eid}{064052} (\bibinfo{year}{2022}), \eprint{2205.01705}.

\bibitem[{\citenamefont{{Kuan} and {Kokkotas}}(2023)}]{Kuan:23}
\bibinfo{author}{\bibfnamefont{H.-J.} \bibnamefont{{Kuan}}} \bibnamefont{and} \bibinfo{author}{\bibfnamefont{K.~D.} \bibnamefont{{Kokkotas}}}, \bibinfo{journal}{\prd} \textbf{\bibinfo{volume}{108}}, \bibinfo{eid}{063026} (\bibinfo{year}{2023}), \eprint{2309.04622}.

\bibitem[{\citenamefont{{Yu} et~al.}(2024)\citenamefont{{Yu}, {Arras}, and {Weinberg}}}]{Yu:24a}
\bibinfo{author}{\bibfnamefont{H.}~\bibnamefont{{Yu}}}, \bibinfo{author}{\bibfnamefont{P.}~\bibnamefont{{Arras}}}, \bibnamefont{and} \bibinfo{author}{\bibfnamefont{N.~N.} \bibnamefont{{Weinberg}}}, \bibinfo{journal}{\prd} \textbf{\bibinfo{volume}{110}}, \bibinfo{eid}{024039} (\bibinfo{year}{2024}).

\bibitem[{\citenamefont{{Abbott} et~al.}(2017)\citenamefont{{Abbott}, {Abbott}, {Abbott}, {Acernese}, {Ackley}, {Adams}, {Adams}, {Addesso}, {Adhikari}, {Adya} et~al.}}]{gw170817}
\bibinfo{author}{\bibfnamefont{B.~P.} \bibnamefont{{Abbott}}}, \bibinfo{author}{\bibfnamefont{R.}~\bibnamefont{{Abbott}}}, \bibinfo{author}{\bibfnamefont{T.~D.} \bibnamefont{{Abbott}}}, \bibinfo{author}{\bibfnamefont{F.}~\bibnamefont{{Acernese}}}, \bibinfo{author}{\bibfnamefont{K.}~\bibnamefont{{Ackley}}}, \bibinfo{author}{\bibfnamefont{C.}~\bibnamefont{{Adams}}}, \bibinfo{author}{\bibfnamefont{T.}~\bibnamefont{{Adams}}}, \bibinfo{author}{\bibfnamefont{P.}~\bibnamefont{{Addesso}}}, \bibinfo{author}{\bibfnamefont{R.~X.} \bibnamefont{{Adhikari}}}, \bibinfo{author}{\bibfnamefont{V.~B.} \bibnamefont{{Adya}}}, \bibnamefont{et~al.}, \bibinfo{journal}{\prl} \textbf{\bibinfo{volume}{119}}, \bibinfo{eid}{161101} (\bibinfo{year}{2017}), \eprint{1710.05832}.

\bibitem[{\citenamefont{Abbott et~al.}(2019)\citenamefont{Abbott, Abbott, Abbott, Acernese, Ackley, Adams, Adams, Addesso, Adhikari, Adya et~al.}}]{gw170817_improve}
\bibinfo{author}{\bibfnamefont{B.~P.} \bibnamefont{Abbott}}, \bibinfo{author}{\bibfnamefont{R.}~\bibnamefont{Abbott}}, \bibinfo{author}{\bibfnamefont{T.~D.} \bibnamefont{Abbott}}, \bibinfo{author}{\bibfnamefont{F.}~\bibnamefont{Acernese}}, \bibinfo{author}{\bibfnamefont{K.}~\bibnamefont{Ackley}}, \bibinfo{author}{\bibfnamefont{C.}~\bibnamefont{Adams}}, \bibinfo{author}{\bibfnamefont{T.}~\bibnamefont{Adams}}, \bibinfo{author}{\bibfnamefont{P.}~\bibnamefont{Addesso}}, \bibinfo{author}{\bibfnamefont{R.~X.} \bibnamefont{Adhikari}}, \bibinfo{author}{\bibfnamefont{V.~B.} \bibnamefont{Adya}}, \bibnamefont{et~al.} (\bibinfo{collaboration}{LIGO Scientific Collaboration and Virgo Collaboration}), \bibinfo{journal}{Phys. Rev. X} \textbf{\bibinfo{volume}{9}}, \bibinfo{pages}{011001} (\bibinfo{year}{2019}), \urlprefix\url{https://link.aps.org/doi/10.1103/PhysRevX.9.011001}.

\bibitem[{\citenamefont{Abbott et~al.}(2018)\citenamefont{Abbott, Abbott, Abbott, Acernese, Ackley, Adams, Adams, Addesso, Adhikari, Adya et~al.}}]{gw170817_eos}
\bibinfo{author}{\bibfnamefont{B.~P.} \bibnamefont{Abbott}}, \bibinfo{author}{\bibfnamefont{R.}~\bibnamefont{Abbott}}, \bibinfo{author}{\bibfnamefont{T.~D.} \bibnamefont{Abbott}}, \bibinfo{author}{\bibfnamefont{F.}~\bibnamefont{Acernese}}, \bibinfo{author}{\bibfnamefont{K.}~\bibnamefont{Ackley}}, \bibinfo{author}{\bibfnamefont{C.}~\bibnamefont{Adams}}, \bibinfo{author}{\bibfnamefont{T.}~\bibnamefont{Adams}}, \bibinfo{author}{\bibfnamefont{P.}~\bibnamefont{Addesso}}, \bibinfo{author}{\bibfnamefont{R.~X.} \bibnamefont{Adhikari}}, \bibinfo{author}{\bibfnamefont{V.~B.} \bibnamefont{Adya}}, \bibnamefont{et~al.} (\bibinfo{collaboration}{The LIGO Scientific Collaboration and the Virgo Collaboration}), \bibinfo{journal}{Phys. Rev. Lett.} \textbf{\bibinfo{volume}{121}}, \bibinfo{pages}{161101} (\bibinfo{year}{2018}), \urlprefix\url{https://link.aps.org/doi/10.1103/PhysRevLett.121.161101}.

\bibitem[{\citenamefont{{Abbott} et~al.}(2020)\citenamefont{{Abbott}, {Abbott}, {Abbott}, {Abraham}, {Acernese}, {Ackley}, {Adams}, {Adya}, {Affeldt}, {Agathos} et~al.}}]{virgo_eos}
\bibinfo{author}{\bibfnamefont{B.~P.} \bibnamefont{{Abbott}}}, \bibinfo{author}{\bibfnamefont{R.}~\bibnamefont{{Abbott}}}, \bibinfo{author}{\bibfnamefont{T.~D.} \bibnamefont{{Abbott}}}, \bibinfo{author}{\bibfnamefont{S.}~\bibnamefont{{Abraham}}}, \bibinfo{author}{\bibfnamefont{F.}~\bibnamefont{{Acernese}}}, \bibinfo{author}{\bibfnamefont{K.}~\bibnamefont{{Ackley}}}, \bibinfo{author}{\bibfnamefont{C.}~\bibnamefont{{Adams}}}, \bibinfo{author}{\bibfnamefont{V.~B.} \bibnamefont{{Adya}}}, \bibinfo{author}{\bibfnamefont{C.}~\bibnamefont{{Affeldt}}}, \bibinfo{author}{\bibfnamefont{M.}~\bibnamefont{{Agathos}}}, \bibnamefont{et~al.}, \bibinfo{journal}{Classical and Quantum Gravity} \textbf{\bibinfo{volume}{37}}, \bibinfo{eid}{045006} (\bibinfo{year}{2020}), \eprint{1908.01012}.

\bibitem[{\citenamefont{{LIGO Scientific Collaboration} et~al.}(2015)\citenamefont{{LIGO Scientific Collaboration}, {Aasi}, {Abbott}, {Abbott}, {Abbott}, {Abernathy}, {Ackley}, {Adams}, {Adams}, {Addesso} et~al.}}]{aligo}
\bibinfo{author}{\bibnamefont{{LIGO Scientific Collaboration}}}, \bibinfo{author}{\bibfnamefont{J.}~\bibnamefont{{Aasi}}}, \bibinfo{author}{\bibfnamefont{B.~P.} \bibnamefont{{Abbott}}}, \bibinfo{author}{\bibfnamefont{R.}~\bibnamefont{{Abbott}}}, \bibinfo{author}{\bibfnamefont{T.}~\bibnamefont{{Abbott}}}, \bibinfo{author}{\bibfnamefont{M.~R.} \bibnamefont{{Abernathy}}}, \bibinfo{author}{\bibfnamefont{K.}~\bibnamefont{{Ackley}}}, \bibinfo{author}{\bibfnamefont{C.}~\bibnamefont{{Adams}}}, \bibinfo{author}{\bibfnamefont{T.}~\bibnamefont{{Adams}}}, \bibinfo{author}{\bibfnamefont{P.}~\bibnamefont{{Addesso}}}, \bibnamefont{et~al.}, \bibinfo{journal}{Classical and Quantum Gravity} \textbf{\bibinfo{volume}{32}}, \bibinfo{eid}{074001} (\bibinfo{year}{2015}), \eprint{1411.4547}.

\bibitem[{\citenamefont{{Acernese} et~al.}(2015)\citenamefont{{Acernese}, {Agathos}, {Agatsuma}, {Aisa}, {Allemandou}, {Allocca}, {Amarni}, {Astone}, {Balestri}, {Ballardin} et~al.}}]{virgo}
\bibinfo{author}{\bibfnamefont{F.}~\bibnamefont{{Acernese}}}, \bibinfo{author}{\bibfnamefont{M.}~\bibnamefont{{Agathos}}}, \bibinfo{author}{\bibfnamefont{K.}~\bibnamefont{{Agatsuma}}}, \bibinfo{author}{\bibfnamefont{D.}~\bibnamefont{{Aisa}}}, \bibinfo{author}{\bibfnamefont{N.}~\bibnamefont{{Allemandou}}}, \bibinfo{author}{\bibfnamefont{A.}~\bibnamefont{{Allocca}}}, \bibinfo{author}{\bibfnamefont{J.}~\bibnamefont{{Amarni}}}, \bibinfo{author}{\bibfnamefont{P.}~\bibnamefont{{Astone}}}, \bibinfo{author}{\bibfnamefont{G.}~\bibnamefont{{Balestri}}}, \bibinfo{author}{\bibfnamefont{G.}~\bibnamefont{{Ballardin}}}, \bibnamefont{et~al.}, \bibinfo{journal}{Classical and Quantum Gravity} \textbf{\bibinfo{volume}{32}}, \bibinfo{eid}{024001} (\bibinfo{year}{2015}), \eprint{1408.3978}.

\bibitem[{\citenamefont{{Somiya}}(2012)}]{kagra}
\bibinfo{author}{\bibfnamefont{K.}~\bibnamefont{{Somiya}}}, \bibinfo{journal}{Classical and Quantum Gravity} \textbf{\bibinfo{volume}{29}}, \bibinfo{eid}{124007} (\bibinfo{year}{2012}), \eprint{1111.7185}.

\bibitem[{\citenamefont{{Reitze} et~al.}(2019)\citenamefont{{Reitze}, {Adhikari}, {Ballmer}, {Barish}, {Barsotti}, {Billingsley}, {Brown}, {Chen}, {Coyne}, {Eisenstein} et~al.}}]{cosmic_explorer}
\bibinfo{author}{\bibfnamefont{D.}~\bibnamefont{{Reitze}}}, \bibinfo{author}{\bibfnamefont{R.~X.} \bibnamefont{{Adhikari}}}, \bibinfo{author}{\bibfnamefont{S.}~\bibnamefont{{Ballmer}}}, \bibinfo{author}{\bibfnamefont{B.}~\bibnamefont{{Barish}}}, \bibinfo{author}{\bibfnamefont{L.}~\bibnamefont{{Barsotti}}}, \bibinfo{author}{\bibfnamefont{G.}~\bibnamefont{{Billingsley}}}, \bibinfo{author}{\bibfnamefont{D.~A.} \bibnamefont{{Brown}}}, \bibinfo{author}{\bibfnamefont{Y.}~\bibnamefont{{Chen}}}, \bibinfo{author}{\bibfnamefont{D.}~\bibnamefont{{Coyne}}}, \bibinfo{author}{\bibfnamefont{R.}~\bibnamefont{{Eisenstein}}}, \bibnamefont{et~al.}, in \emph{\bibinfo{booktitle}{Bulletin of the American Astronomical Society}} (\bibinfo{year}{2019}), vol.~\bibinfo{volume}{51}, p.~\bibinfo{pages}{35}, \eprint{1907.04833}.

\bibitem[{\citenamefont{{Punturo} et~al.}(2010)\citenamefont{{Punturo}, {Abernathy}, {Acernese}, {Allen}, {Andersson}, {Arun}, {Barone}, {Barr}, {Barsuglia}, {Beker} et~al.}}]{einstein_telescope}
\bibinfo{author}{\bibfnamefont{M.}~\bibnamefont{{Punturo}}}, \bibinfo{author}{\bibfnamefont{M.}~\bibnamefont{{Abernathy}}}, \bibinfo{author}{\bibfnamefont{F.}~\bibnamefont{{Acernese}}}, \bibinfo{author}{\bibfnamefont{B.}~\bibnamefont{{Allen}}}, \bibinfo{author}{\bibfnamefont{N.}~\bibnamefont{{Andersson}}}, \bibinfo{author}{\bibfnamefont{K.}~\bibnamefont{{Arun}}}, \bibinfo{author}{\bibfnamefont{F.}~\bibnamefont{{Barone}}}, \bibinfo{author}{\bibfnamefont{B.}~\bibnamefont{{Barr}}}, \bibinfo{author}{\bibfnamefont{M.}~\bibnamefont{{Barsuglia}}}, \bibinfo{author}{\bibfnamefont{M.}~\bibnamefont{{Beker}}}, \bibnamefont{et~al.}, \bibinfo{journal}{Classical and Quantum Gravity} \textbf{\bibinfo{volume}{27}}, \bibinfo{eid}{194002} (\bibinfo{year}{2010}).

\bibitem[{\citenamefont{{Abbott} et~al.}(2018)\citenamefont{{Abbott}, {Abbott}, {Abbott}, {Abernathy}, {Acernese}, {Ackley}, {Adams}, {Adams}, {Addesso}, {Adhikari} et~al.}}]{prospect18}
\bibinfo{author}{\bibfnamefont{B.~P.} \bibnamefont{{Abbott}}}, \bibinfo{author}{\bibfnamefont{R.}~\bibnamefont{{Abbott}}}, \bibinfo{author}{\bibfnamefont{T.~D.} \bibnamefont{{Abbott}}}, \bibinfo{author}{\bibfnamefont{M.~R.} \bibnamefont{{Abernathy}}}, \bibinfo{author}{\bibfnamefont{F.}~\bibnamefont{{Acernese}}}, \bibinfo{author}{\bibfnamefont{K.}~\bibnamefont{{Ackley}}}, \bibinfo{author}{\bibfnamefont{C.}~\bibnamefont{{Adams}}}, \bibinfo{author}{\bibfnamefont{T.}~\bibnamefont{{Adams}}}, \bibinfo{author}{\bibfnamefont{P.}~\bibnamefont{{Addesso}}}, \bibinfo{author}{\bibfnamefont{R.~X.} \bibnamefont{{Adhikari}}}, \bibnamefont{et~al.}, \bibinfo{journal}{Living Reviews in Relativity} \textbf{\bibinfo{volume}{21}}, \bibinfo{eid}{3} (\bibinfo{year}{2018}), \eprint{1304.0670}.

\bibitem[{\citenamefont{Chan et~al.}(2018)\citenamefont{Chan, Messenger, Heng, and Hendry}}]{chan18}
\bibinfo{author}{\bibfnamefont{M.~L.} \bibnamefont{Chan}}, \bibinfo{author}{\bibfnamefont{C.}~\bibnamefont{Messenger}}, \bibinfo{author}{\bibfnamefont{I.~S.} \bibnamefont{Heng}}, \bibnamefont{and} \bibinfo{author}{\bibfnamefont{M.}~\bibnamefont{Hendry}}, \bibinfo{journal}{Phys. Rev. D} \textbf{\bibinfo{volume}{97}}, \bibinfo{pages}{123014} (\bibinfo{year}{2018}), \urlprefix\url{https://link.aps.org/doi/10.1103/PhysRevD.97.123014}.

\bibitem[{\citenamefont{Lenon et~al.}(2021)\citenamefont{Lenon, Brown, and Nitz}}]{lenon21}
\bibinfo{author}{\bibfnamefont{A.~K.} \bibnamefont{Lenon}}, \bibinfo{author}{\bibfnamefont{D.~A.} \bibnamefont{Brown}}, \bibnamefont{and} \bibinfo{author}{\bibfnamefont{A.~H.} \bibnamefont{Nitz}}, \bibinfo{journal}{Phys. Rev. D} \textbf{\bibinfo{volume}{104}}, \bibinfo{pages}{063011} (\bibinfo{year}{2021}), \urlprefix\url{https://link.aps.org/doi/10.1103/PhysRevD.104.063011}.

\bibitem[{\citenamefont{{Mukhopadhyay} et~al.}(2024)\citenamefont{{Mukhopadhyay}, {Kimura}, and {Murase}}}]{mukhopadhyay24}
\bibinfo{author}{\bibfnamefont{M.}~\bibnamefont{{Mukhopadhyay}}}, \bibinfo{author}{\bibfnamefont{S.~S.} \bibnamefont{{Kimura}}}, \bibnamefont{and} \bibinfo{author}{\bibfnamefont{K.}~\bibnamefont{{Murase}}}, \bibinfo{journal}{\prd} \textbf{\bibinfo{volume}{109}}, \bibinfo{eid}{043053} (\bibinfo{year}{2024}), \eprint{2310.16875}.

\bibitem[{\citenamefont{{Yu} and {Weinberg}}(2017{\natexlab{a}})}]{Yu:17a}
\bibinfo{author}{\bibfnamefont{H.}~\bibnamefont{{Yu}}} \bibnamefont{and} \bibinfo{author}{\bibfnamefont{N.~N.} \bibnamefont{{Weinberg}}}, \bibinfo{journal}{\mnras} \textbf{\bibinfo{volume}{464}}, \bibinfo{pages}{2622} (\bibinfo{year}{2017}{\natexlab{a}}), \eprint{1610.00745}.

\bibitem[{\citenamefont{{Yu} and {Weinberg}}(2017{\natexlab{b}})}]{Yu:17b}
\bibinfo{author}{\bibfnamefont{H.}~\bibnamefont{{Yu}}} \bibnamefont{and} \bibinfo{author}{\bibfnamefont{N.~N.} \bibnamefont{{Weinberg}}}, \bibinfo{journal}{\mnras} \textbf{\bibinfo{volume}{470}}, \bibinfo{pages}{350} (\bibinfo{year}{2017}{\natexlab{b}}), \eprint{1705.04700}.

\bibitem[{\citenamefont{{Kuan} et~al.}(2021{\natexlab{a}})\citenamefont{{Kuan}, {Suvorov}, and {Kokkotas}}}]{Kuan:21}
\bibinfo{author}{\bibfnamefont{H.-J.} \bibnamefont{{Kuan}}}, \bibinfo{author}{\bibfnamefont{A.~G.} \bibnamefont{{Suvorov}}}, \bibnamefont{and} \bibinfo{author}{\bibfnamefont{K.~D.} \bibnamefont{{Kokkotas}}}, \bibinfo{journal}{\mnras} \textbf{\bibinfo{volume}{506}}, \bibinfo{pages}{2985} (\bibinfo{year}{2021}{\natexlab{a}}), \eprint{2106.16123}.

\bibitem[{\citenamefont{{Kuan} et~al.}(2021{\natexlab{b}})\citenamefont{{Kuan}, {Suvorov}, and {Kokkotas}}}]{Kuan:21b}
\bibinfo{author}{\bibfnamefont{H.-J.} \bibnamefont{{Kuan}}}, \bibinfo{author}{\bibfnamefont{A.~G.} \bibnamefont{{Suvorov}}}, \bibnamefont{and} \bibinfo{author}{\bibfnamefont{K.~D.} \bibnamefont{{Kokkotas}}}, \bibinfo{journal}{\mnras} \textbf{\bibinfo{volume}{508}}, \bibinfo{pages}{1732} (\bibinfo{year}{2021}{\natexlab{b}}), \eprint{2107.00533}.

\bibitem[{\citenamefont{{Van Hoolst}}(1994)}]{VanHoolst:94}
\bibinfo{author}{\bibfnamefont{T.}~\bibnamefont{{Van Hoolst}}}, \bibinfo{journal}{\aap} \textbf{\bibinfo{volume}{286}}, \bibinfo{pages}{879} (\bibinfo{year}{1994}).

\bibitem[{\citenamefont{{Schenk} et~al.}(2002)\citenamefont{{Schenk}, {Arras}, {Flanagan}, {Teukolsky}, and {Wasserman}}}]{Schenk:02}
\bibinfo{author}{\bibfnamefont{A.~K.} \bibnamefont{{Schenk}}}, \bibinfo{author}{\bibfnamefont{P.}~\bibnamefont{{Arras}}}, \bibinfo{author}{\bibfnamefont{{\'E}.~{\'E}.} \bibnamefont{{Flanagan}}}, \bibinfo{author}{\bibfnamefont{S.~A.} \bibnamefont{{Teukolsky}}}, \bibnamefont{and} \bibinfo{author}{\bibfnamefont{I.}~\bibnamefont{{Wasserman}}}, \bibinfo{journal}{\prd} \textbf{\bibinfo{volume}{65}}, \bibinfo{eid}{024001} (\bibinfo{year}{2002}), \eprint{gr-qc/0101092}.

\bibitem[{\citenamefont{{Yu} et~al.}(2023)\citenamefont{{Yu}, {Weinberg}, {Arras}, {Kwon}, and {Venumadhav}}}]{Yu:23a}
\bibinfo{author}{\bibfnamefont{H.}~\bibnamefont{{Yu}}}, \bibinfo{author}{\bibfnamefont{N.~N.} \bibnamefont{{Weinberg}}}, \bibinfo{author}{\bibfnamefont{P.}~\bibnamefont{{Arras}}}, \bibinfo{author}{\bibfnamefont{J.}~\bibnamefont{{Kwon}}}, \bibnamefont{and} \bibinfo{author}{\bibfnamefont{T.}~\bibnamefont{{Venumadhav}}}, \bibinfo{journal}{\mnras} \textbf{\bibinfo{volume}{519}}, \bibinfo{pages}{4325} (\bibinfo{year}{2023}), \eprint{2211.07002}.

\bibitem[{\citenamefont{Landau and Lifshitz}(1982)}]{Landau:82}
\bibinfo{author}{\bibfnamefont{L.}~\bibnamefont{Landau}} \bibnamefont{and} \bibinfo{author}{\bibfnamefont{E.}~\bibnamefont{Lifshitz}}, \emph{\bibinfo{title}{Mechanics: Volume 1}}, \bibinfo{number}{v. 1} (\bibinfo{publisher}{Elsevier Science}, \bibinfo{year}{1982}), ISBN \bibinfo{isbn}{9780080503479}, \urlprefix\url{https://books.google.com/books?id=bE-9tUH2J2wC}.

\bibitem[{\citenamefont{{Yu} et~al.}(2021)\citenamefont{{Yu}, {Weinberg}, and {Arras}}}]{Yu:21}
\bibinfo{author}{\bibfnamefont{H.}~\bibnamefont{{Yu}}}, \bibinfo{author}{\bibfnamefont{N.~N.} \bibnamefont{{Weinberg}}}, \bibnamefont{and} \bibinfo{author}{\bibfnamefont{P.}~\bibnamefont{{Arras}}}, \bibinfo{journal}{\apj} \textbf{\bibinfo{volume}{917}}, \bibinfo{eid}{31} (\bibinfo{year}{2021}), \eprint{2104.04929}.

\bibitem[{\citenamefont{{Weinberg} et~al.}(2012{\natexlab{a}})\citenamefont{{Weinberg}, {Arras}, {Quataert}, and {Burkart}}}]{w12}
\bibinfo{author}{\bibfnamefont{N.~N.} \bibnamefont{{Weinberg}}}, \bibinfo{author}{\bibfnamefont{P.}~\bibnamefont{{Arras}}}, \bibinfo{author}{\bibfnamefont{E.}~\bibnamefont{{Quataert}}}, \bibnamefont{and} \bibinfo{author}{\bibfnamefont{J.}~\bibnamefont{{Burkart}}}, \bibinfo{journal}{\apj} \textbf{\bibinfo{volume}{751}}, \bibinfo{eid}{136} (\bibinfo{year}{2012}{\natexlab{a}}), \eprint{1107.0946}.

\bibitem[{\citenamefont{{Venumadhav} et~al.}(2014{\natexlab{a}})\citenamefont{{Venumadhav}, {Zimmerman}, and {Hirata}}}]{vzh}
\bibinfo{author}{\bibfnamefont{T.}~\bibnamefont{{Venumadhav}}}, \bibinfo{author}{\bibfnamefont{A.}~\bibnamefont{{Zimmerman}}}, \bibnamefont{and} \bibinfo{author}{\bibfnamefont{C.~M.} \bibnamefont{{Hirata}}}, \bibinfo{journal}{\apj} \textbf{\bibinfo{volume}{781}}, \bibinfo{eid}{23} (\bibinfo{year}{2014}{\natexlab{a}}), \eprint{1307.2890}.

\bibitem[{\citenamefont{{Weinberg}}(2016)}]{w16}
\bibinfo{author}{\bibfnamefont{N.~N.} \bibnamefont{{Weinberg}}}, \bibinfo{journal}{\apj} \textbf{\bibinfo{volume}{819}}, \bibinfo{eid}{109} (\bibinfo{year}{2016}), \eprint{1509.06975}.

\bibitem[{\citenamefont{{Kwon} et~al.}(2025)\citenamefont{{Kwon}, {Yu}, and {Venumadhav}}}]{nlrl_prd}
\bibinfo{author}{\bibfnamefont{K.~J.} \bibnamefont{{Kwon}}}, \bibinfo{author}{\bibfnamefont{H.}~\bibnamefont{{Yu}}}, \bibnamefont{and} \bibinfo{author}{\bibfnamefont{T.}~\bibnamefont{{Venumadhav}}}, \bibinfo{journal}{arXiv e-prints} \bibinfo{eid}{arXiv:2503.11837} (\bibinfo{year}{2025}), \eprint{2503.11837}.

\bibitem[{\citenamefont{{Fuller} and {Lai}}(2012)}]{Fuller:12a}
\bibinfo{author}{\bibfnamefont{J.}~\bibnamefont{{Fuller}}} \bibnamefont{and} \bibinfo{author}{\bibfnamefont{D.}~\bibnamefont{{Lai}}}, \bibinfo{journal}{\mnras} \textbf{\bibinfo{volume}{421}}, \bibinfo{pages}{426} (\bibinfo{year}{2012}), \eprint{1108.4910}.

\bibitem[{\citenamefont{{Burkart} et~al.}(2013)\citenamefont{{Burkart}, {Quataert}, {Arras}, and {Weinberg}}}]{Burkart:13}
\bibinfo{author}{\bibfnamefont{J.}~\bibnamefont{{Burkart}}}, \bibinfo{author}{\bibfnamefont{E.}~\bibnamefont{{Quataert}}}, \bibinfo{author}{\bibfnamefont{P.}~\bibnamefont{{Arras}}}, \bibnamefont{and} \bibinfo{author}{\bibfnamefont{N.~N.} \bibnamefont{{Weinberg}}}, \bibinfo{journal}{\mnras} \textbf{\bibinfo{volume}{433}}, \bibinfo{pages}{332} (\bibinfo{year}{2013}), \eprint{1211.1393}.

\bibitem[{\citenamefont{{Burkart} et~al.}(2014)\citenamefont{{Burkart}, {Quataert}, and {Arras}}}]{burkart14}
\bibinfo{author}{\bibfnamefont{J.}~\bibnamefont{{Burkart}}}, \bibinfo{author}{\bibfnamefont{E.}~\bibnamefont{{Quataert}}}, \bibnamefont{and} \bibinfo{author}{\bibfnamefont{P.}~\bibnamefont{{Arras}}}, \bibinfo{journal}{\mnras} \textbf{\bibinfo{volume}{443}}, \bibinfo{pages}{2957} (\bibinfo{year}{2014}), \eprint{1312.4966}.

\bibitem[{\citenamefont{{Yu} et~al.}(2020)\citenamefont{{Yu}, {Weinberg}, and {Fuller}}}]{Yu:20a}
\bibinfo{author}{\bibfnamefont{H.}~\bibnamefont{{Yu}}}, \bibinfo{author}{\bibfnamefont{N.~N.} \bibnamefont{{Weinberg}}}, \bibnamefont{and} \bibinfo{author}{\bibfnamefont{J.}~\bibnamefont{{Fuller}}}, \bibinfo{journal}{\mnras} \textbf{\bibinfo{volume}{496}}, \bibinfo{pages}{5482} (\bibinfo{year}{2020}), \eprint{2005.03058}.

\bibitem[{\citenamefont{Chabanat et~al.}(1998)\citenamefont{Chabanat, Bonche, Haensel, Meyer, and Schaeffer}}]{chabanat98}
\bibinfo{author}{\bibfnamefont{E.}~\bibnamefont{Chabanat}}, \bibinfo{author}{\bibfnamefont{P.}~\bibnamefont{Bonche}}, \bibinfo{author}{\bibfnamefont{P.}~\bibnamefont{Haensel}}, \bibinfo{author}{\bibfnamefont{J.}~\bibnamefont{Meyer}}, \bibnamefont{and} \bibinfo{author}{\bibfnamefont{R.}~\bibnamefont{Schaeffer}}, \bibinfo{journal}{Nuclear Physics A} \textbf{\bibinfo{volume}{635}}, \bibinfo{pages}{231} (\bibinfo{year}{1998}), ISSN \bibinfo{issn}{0375-9474}, \urlprefix\url{https://www.sciencedirect.com/science/article/pii/S0375947498001808}.

\bibitem[{\citenamefont{{Hotokezaka} et~al.}(2013)\citenamefont{{Hotokezaka}, {Kyutoku}, and {Shibata}}}]{hotokezaka13}
\bibinfo{author}{\bibfnamefont{K.}~\bibnamefont{{Hotokezaka}}}, \bibinfo{author}{\bibfnamefont{K.}~\bibnamefont{{Kyutoku}}}, \bibnamefont{and} \bibinfo{author}{\bibfnamefont{M.}~\bibnamefont{{Shibata}}}, \bibinfo{journal}{\prd} \textbf{\bibinfo{volume}{87}}, \bibinfo{eid}{044001} (\bibinfo{year}{2013}), \eprint{1301.3555}.

\bibitem[{\citenamefont{{Hotokezaka} et~al.}(2015)\citenamefont{{Hotokezaka}, {Kyutoku}, {Okawa}, and {Shibata}}}]{Hotokezaka:15}
\bibinfo{author}{\bibfnamefont{K.}~\bibnamefont{{Hotokezaka}}}, \bibinfo{author}{\bibfnamefont{K.}~\bibnamefont{{Kyutoku}}}, \bibinfo{author}{\bibfnamefont{H.}~\bibnamefont{{Okawa}}}, \bibnamefont{and} \bibinfo{author}{\bibfnamefont{M.}~\bibnamefont{{Shibata}}}, \bibinfo{journal}{\prd} \textbf{\bibinfo{volume}{91}}, \bibinfo{eid}{064060} (\bibinfo{year}{2015}), \eprint{1502.03457}.

\bibitem[{\citenamefont{{Dietrich} et~al.}(2018)\citenamefont{{Dietrich}, {Bernuzzi}, {Br{\"u}gmann}, {Ujevic}, and {Tichy}}}]{Dietrich:18}
\bibinfo{author}{\bibfnamefont{T.}~\bibnamefont{{Dietrich}}}, \bibinfo{author}{\bibfnamefont{S.}~\bibnamefont{{Bernuzzi}}}, \bibinfo{author}{\bibfnamefont{B.}~\bibnamefont{{Br{\"u}gmann}}}, \bibinfo{author}{\bibfnamefont{M.}~\bibnamefont{{Ujevic}}}, \bibnamefont{and} \bibinfo{author}{\bibfnamefont{W.}~\bibnamefont{{Tichy}}}, \bibinfo{journal}{\prd} \textbf{\bibinfo{volume}{97}}, \bibinfo{eid}{064002} (\bibinfo{year}{2018}), \eprint{1712.02992}.

\bibitem[{\citenamefont{{Foucart} et~al.}(2019)\citenamefont{{Foucart}, {Duez}, {Hinderer}, {Caro}, {Williamson}, {Boyle}, {Buonanno}, {Haas}, {Hemberger}, {Kidder} et~al.}}]{Foucart:19}
\bibinfo{author}{\bibfnamefont{F.}~\bibnamefont{{Foucart}}}, \bibinfo{author}{\bibfnamefont{M.~D.} \bibnamefont{{Duez}}}, \bibinfo{author}{\bibfnamefont{T.}~\bibnamefont{{Hinderer}}}, \bibinfo{author}{\bibfnamefont{J.}~\bibnamefont{{Caro}}}, \bibinfo{author}{\bibfnamefont{A.~R.} \bibnamefont{{Williamson}}}, \bibinfo{author}{\bibfnamefont{M.}~\bibnamefont{{Boyle}}}, \bibinfo{author}{\bibfnamefont{A.}~\bibnamefont{{Buonanno}}}, \bibinfo{author}{\bibfnamefont{R.}~\bibnamefont{{Haas}}}, \bibinfo{author}{\bibfnamefont{D.~A.} \bibnamefont{{Hemberger}}}, \bibinfo{author}{\bibfnamefont{L.~E.} \bibnamefont{{Kidder}}}, \bibnamefont{et~al.}, \bibinfo{journal}{\prd} \textbf{\bibinfo{volume}{99}}, \bibinfo{eid}{044008} (\bibinfo{year}{2019}), \eprint{1812.06988}.

\bibitem[{\citenamefont{{Ho} and {Lai}}(1999)}]{Ho:99}
\bibinfo{author}{\bibfnamefont{W.~C.~G.} \bibnamefont{{Ho}}} \bibnamefont{and} \bibinfo{author}{\bibfnamefont{D.}~\bibnamefont{{Lai}}}, \bibinfo{journal}{\mnras} \textbf{\bibinfo{volume}{308}}, \bibinfo{pages}{153} (\bibinfo{year}{1999}), \eprint{astro-ph/9812116}.

\bibitem[{\citenamefont{{Xu} and {Lai}}(2017)}]{Xu:17}
\bibinfo{author}{\bibfnamefont{W.}~\bibnamefont{{Xu}}} \bibnamefont{and} \bibinfo{author}{\bibfnamefont{D.}~\bibnamefont{{Lai}}}, \bibinfo{journal}{\prd} \textbf{\bibinfo{volume}{96}}, \bibinfo{eid}{083005} (\bibinfo{year}{2017}), \eprint{1708.01839}.

\bibitem[{\citenamefont{{Yakovlev} et~al.}(1999)\citenamefont{{Yakovlev}, {Levenfish}, and {Shibanov}}}]{Yakovlev:99}
\bibinfo{author}{\bibfnamefont{D.~G.} \bibnamefont{{Yakovlev}}}, \bibinfo{author}{\bibfnamefont{K.~P.} \bibnamefont{{Levenfish}}}, \bibnamefont{and} \bibinfo{author}{\bibfnamefont{Y.~A.} \bibnamefont{{Shibanov}}}, \bibinfo{journal}{Physics Uspekhi} \textbf{\bibinfo{volume}{42}}, \bibinfo{pages}{737} (\bibinfo{year}{1999}), \eprint{astro-ph/9906456}.

\bibitem[{\citenamefont{{Kantor} and {Gusakov}}(2014)}]{Kantor:14}
\bibinfo{author}{\bibfnamefont{E.~M.} \bibnamefont{{Kantor}}} \bibnamefont{and} \bibinfo{author}{\bibfnamefont{M.~E.} \bibnamefont{{Gusakov}}}, \bibinfo{journal}{\mnras} \textbf{\bibinfo{volume}{442}}, \bibinfo{pages}{L90} (\bibinfo{year}{2014}), \eprint{1404.6768}.

\bibitem[{\citenamefont{{Passamonti} et~al.}(2016)\citenamefont{{Passamonti}, {Andersson}, and {Ho}}}]{Passamonti:16}
\bibinfo{author}{\bibfnamefont{A.}~\bibnamefont{{Passamonti}}}, \bibinfo{author}{\bibfnamefont{N.}~\bibnamefont{{Andersson}}}, \bibnamefont{and} \bibinfo{author}{\bibfnamefont{W.~C.~G.} \bibnamefont{{Ho}}}, \bibinfo{journal}{\mnras} \textbf{\bibinfo{volume}{455}}, \bibinfo{pages}{1489} (\bibinfo{year}{2016}), \eprint{1504.07470}.

\bibitem[{\citenamefont{{Tsang} et~al.}(2012)\citenamefont{{Tsang}, {Read}, {Hinderer}, {Piro}, and {Bondarescu}}}]{Tsang:12}
\bibinfo{author}{\bibfnamefont{D.}~\bibnamefont{{Tsang}}}, \bibinfo{author}{\bibfnamefont{J.~S.} \bibnamefont{{Read}}}, \bibinfo{author}{\bibfnamefont{T.}~\bibnamefont{{Hinderer}}}, \bibinfo{author}{\bibfnamefont{A.~L.} \bibnamefont{{Piro}}}, \bibnamefont{and} \bibinfo{author}{\bibfnamefont{R.}~\bibnamefont{{Bondarescu}}}, \bibinfo{journal}{\prl} \textbf{\bibinfo{volume}{108}}, \bibinfo{eid}{011102} (\bibinfo{year}{2012}), \eprint{1110.0467}.

\bibitem[{\citenamefont{{Passamonti} et~al.}(2021)\citenamefont{{Passamonti}, {Andersson}, and {Pnigouras}}}]{Passamonti:21}
\bibinfo{author}{\bibfnamefont{A.}~\bibnamefont{{Passamonti}}}, \bibinfo{author}{\bibfnamefont{N.}~\bibnamefont{{Andersson}}}, \bibnamefont{and} \bibinfo{author}{\bibfnamefont{P.}~\bibnamefont{{Pnigouras}}}, \bibinfo{journal}{\mnras} \textbf{\bibinfo{volume}{504}}, \bibinfo{pages}{1273} (\bibinfo{year}{2021}), \eprint{2012.09637}.

\bibitem[{\citenamefont{{Pan} et~al.}(2020)\citenamefont{{Pan}, {Lyu}, {Bonga}, {Ortiz}, and {Yang}}}]{Pan:20}
\bibinfo{author}{\bibfnamefont{Z.}~\bibnamefont{{Pan}}}, \bibinfo{author}{\bibfnamefont{Z.}~\bibnamefont{{Lyu}}}, \bibinfo{author}{\bibfnamefont{B.}~\bibnamefont{{Bonga}}}, \bibinfo{author}{\bibfnamefont{N.}~\bibnamefont{{Ortiz}}}, \bibnamefont{and} \bibinfo{author}{\bibfnamefont{H.}~\bibnamefont{{Yang}}}, \bibinfo{journal}{\prl} \textbf{\bibinfo{volume}{125}}, \bibinfo{eid}{201102} (\bibinfo{year}{2020}), \eprint{2003.03330}.

\bibitem[{\citenamefont{{Weinberg} et~al.}(2012{\natexlab{b}})\citenamefont{{Weinberg}, {Arras}, {Quataert}, and {Burkart}}}]{Weinberg:12}
\bibinfo{author}{\bibfnamefont{N.~N.} \bibnamefont{{Weinberg}}}, \bibinfo{author}{\bibfnamefont{P.}~\bibnamefont{{Arras}}}, \bibinfo{author}{\bibfnamefont{E.}~\bibnamefont{{Quataert}}}, \bibnamefont{and} \bibinfo{author}{\bibfnamefont{J.}~\bibnamefont{{Burkart}}}, \bibinfo{journal}{\apj} \textbf{\bibinfo{volume}{751}}, \bibinfo{eid}{136} (\bibinfo{year}{2012}{\natexlab{b}}), \eprint{1107.0946}.

\bibitem[{\citenamefont{{Weinberg} et~al.}(2013)\citenamefont{{Weinberg}, {Arras}, and {Burkart}}}]{Weinberg:13}
\bibinfo{author}{\bibfnamefont{N.~N.} \bibnamefont{{Weinberg}}}, \bibinfo{author}{\bibfnamefont{P.}~\bibnamefont{{Arras}}}, \bibnamefont{and} \bibinfo{author}{\bibfnamefont{J.}~\bibnamefont{{Burkart}}}, \bibinfo{journal}{\apj} \textbf{\bibinfo{volume}{769}}, \bibinfo{eid}{121} (\bibinfo{year}{2013}), \eprint{1302.2292}.

\bibitem[{\citenamefont{{Venumadhav} et~al.}(2014{\natexlab{b}})\citenamefont{{Venumadhav}, {Zimmerman}, and {Hirata}}}]{Venumadhav:14}
\bibinfo{author}{\bibfnamefont{T.}~\bibnamefont{{Venumadhav}}}, \bibinfo{author}{\bibfnamefont{A.}~\bibnamefont{{Zimmerman}}}, \bibnamefont{and} \bibinfo{author}{\bibfnamefont{C.~M.} \bibnamefont{{Hirata}}}, \bibinfo{journal}{ApJ} \textbf{\bibinfo{volume}{781}}, \bibinfo{eid}{23} (\bibinfo{year}{2014}{\natexlab{b}}), \eprint{1307.2890}.

\end{thebibliography}

\end{document}